\documentclass{article}
\usepackage{arxiv}

\usepackage[utf8]{inputenc}
\usepackage[T1]{fontenc}

\usepackage{url}
\usepackage{graphicx}
\usepackage[hidelinks]{hyperref}

\usepackage{orcidlink}

\usepackage{biblatex}
\bibliography{bibliography}

\usepackage[group-separator={,}]{siunitx}

\usepackage{subcaption}
\DeclareGraphicsExtensions{.pdf}

\usepackage{glossaries}
\makeglossaries
\newacronym{API}{API}{Application Programming Interface}
\newacronym{CEU}{CEU}{Central European University}
\newacronym{CDR}{CDR}{Call Detail Record}
\newacronym{CSV}{CSV}{Comma Separated Value}
\newacronym{ETL}{ETL}{extract, transform, load}
\newacronym{GIS}{GIS}{Geographic Information System}
\newacronym{GPS}{GPS}{Global Positioning System}
\newacronym{HUF}{HUF}{Hungarian forint} % the currency of Hungary
\newacronym{IMEI}{IMEI}{International Mobile Equipment Identity}
\newacronym{KSH}{KSH}{Központi Statisztikai Hivatal, Hungarian Central Statistical Office}
\newacronym{MTI}{MTI}{Magyar Távirati Iroda, Hungarian news agency}
\newacronym{OSM}{OSM}{OpenStreetMap}
\newacronym{PCA}{PCA}{Principal Component Analysis}
\newacronym{POI}{POI}{Point of interest}
\newacronym{SES}{SES}{Social Economic Status}
\newacronym{SWC}{SWC}{sleep wake cycle}
\newacronym{SIM}{SIM}{Subscriber Identity Module}
\newacronym{TAC}{TAC}{Type Allocation Code}
\newacronym{UMTS}{UMTS}{Universal Mobile Telecommunications System}
\newacronym{WGS84}{WGS 84}{World Geodetic System, also known as EPSG:4326}
\newacronym{WKT}{WKT}{Well-known text}

\usepackage[titletoc,title]{appendix}

\newcommand{\DailyEntWuPearsonR}{-0.8932}
\newcommand{\DailyGyrWuPearsonR}{-0.6873}
\newcommand{\DailyEntDoPearsonR}{-0.85}
\newcommand{\DailyGyrDoPearsonR}{-0.6621}

\begin{document}

\title{Awakening City: Traces of the Circadian Rhythm within the Mobile Phone Network Data}

\author{Gerg\H{o} Pint\'{e}r \orcidlink{0000-0003-4731-3816} \\
John von Neumann Faculty of Informatics, \\
\'{O}buda University, \\
B\'{e}csi \'{u}t 96/B, 1034 Budapest, Hungary \\
\texttt{pinter.gergo@uni-obuda.hu} \And
Imre Felde \orcidlink{0000-0003-4126-2480} \\
John von Neumann Faculty of Informatics,\\
\'{O}buda University,\\
B\'{e}csi \'{u}t 96/B, 1034 Budapest, Hungary \\
\texttt{felde.imre@uni-obuda.hu}
}

\keywords{mobile phone data; call detail records; type allocation code; data analysis; human mobility; urban mobility; social sensing; socioeconomic status; circadian rhythm; sleep wake cycle}

\maketitle{}

\begin{abstract}
In this study, Call Detail Records (CDR), covering Budapest, Hungary has been processed to analyze the circadian rhythm of the subscribers.
An indicator, called wake-up time, is introduced to describe a behavior of a group of subscribers.
It is defined as the time, when the mobile phone activity of a group rises in the morning. Its counterpart is the time, when the activity falls in the evening.
Inhabitant and area-based aggregation are also presented. The former is to consider the people who live in an area, the latter uses the transit activity in an area to describe the behavior of a part of the city. The opening hours of the malls and the nightlife of the party district was used to demonstrate this application, as real-life examples.
The proposed approach was also used to estimate the working hours of the workplaces. The findings are in a good agreement with practice in Hungary, and also support the workplace detection method.
Negative correlation was found between wake-up time and mobility indicators (Entropy, Radius of Gyration): On workdays, people wake up earlier and travel more, on holidays it is quite the contrary.
The wake-up time was evaluated in different socioeconomic classes, using housing prices and mobile phones prices, as well. It was found that lower socioeconomic groups tend to wake up earlier.
\end{abstract}

\section{Introduction}

% general intro
The mobile phone network, during its operation, constantly communicates with the cell phones. This communication can be divided into two categories: (i) the passive, cell-switching communication, that keeps the cell phones ready to use the mobile phone network at any time, and (ii) the active, billed usage of the mobile phone network, including phone calls, text messages or mobile internet usage. The Call Detail Records (\acrshort{CDR}) collect the latter, containing information about the subscriber, the time of the activity and the place (via the cell), where the activity is occurred.

In the last few decades, anonymized \acrshort{CDR} have become a standard information source for analyzing the characteristics of human mobility. The billed activities of the subscribers are recorded, providing information about the whereabouts of the population.
The human mobility analysis, based on this massive information source, is utilized in fields --- among others --- like social sensing, epidemiology, transportation engineering, urban planning or sociology.
Furthermore, human sleep wake cycle (\acrshort{SWC}) is also studied by analyzing mobile phone network data.

% social sensing
\acrshort{CDR} processing is often applied for large social event detection, such as football matches \cite{traag2011social,xavier2012analyzing,mamei2016estimating,pinter2021analyzing}, concerts \cite{marques2018understanding}, sociopolitical events \cite{furletti2017discovering,hiir2020impact} or mass protests \cite{pinter2019activity,rotman2020using}.
% COVID-19
Epidemiology is used to be mentioned as a potential application of human mobility studies, but the COVID-19 pandemic prioritized its applications in Digital Epidemiology, as mobile phone network data can reflect the mobility changes, caused by the imposed restrictions.
Willberg et al. found a considerable decrease of the population presence in the largest cities of Finland, after the lockdown \cite{willberg2021escaping}. Romanillos et al. reported similar results from the Madrid metropolitan area \cite{romanillos2021city}.
Lee et al. examined the mobility changes during the lockdown in England, and found that the mobility of the wealthier subscribers decreased more significantly \cite{do2021association}.
Khataee et al. compared the effect of the social distancing in several countries, using mobility data from Apple phones \cite{khataee2021effects}.
Bushman et al. \cite{bushman2020effectiveness}, Gao et al. \cite{gao2020association}, Hu et al. \cite{hu2021big} and Tokey \cite{tokey2021spatial} also analyzed effects of the stay-at-home distancing on the COVID-19 increase rate, in the US.
Lucchini et al. studied the mobility changes during the pandemic, in four US states \cite{lucchini2021living}.

% commuting
Identifying the home and work locations of a subscriber is a common \cite{pappalardo2015returners,xu2015understanding,jiang2017activity,vanhoof2018assessing,xu2018human,zagatti2018trip,mamei2019evaluating,pappalardo2021evaluation,pinter2021evaluating} and crucial part of the \acrshort{CDR} processing, as these locations fundamentally determine the people's mobility customs. Furthermore, a good portion of the people live their lives in an area, that is determined by only their home and workplace \cite{pappalardo2015returners,jiang2017activity} or their communities \cite{ghahramani2018extracting}.
Using the home and the work locations, the commuting trends can be analyzed \cite{csaji2013exploring,kung2014exploring,zagatti2018trip,mamei2019evaluating,pinter2021evaluating}.

% SES
Analyzing city structure led to the analyses of the socioeconomic structure of the population, as different social classes live in different parts of a city, but \acrshort{CDR}s also have been used to analyze gender and minority segregation \cite{goel2021understanding}.
Xu et al. \cite{xu2018human} used six mobility indicators, housing prices and per capita income in Singapore and Boston to analyze the socioeconomic classes. It was found that the wealthier subscribers tend to travel shorter distances in Singapore, but longer, in Boston.
Barbosa et al. also found significant differences in the average travel distance between the low and high income groups in Brazil \cite{barbosa2020uncovering}.
In a previous work, we have also demonstrated differences in mobility customs between socioeconomic classes \cite{pinter2021evaluating}, in the case of Budapest.
Ucar et al. revealed socioeconomic gap by mobile service consumption \cite{ucar2021news}. Vilella et al. found that education and age play news media consumption patterns, in Chile, using a dataset, that provides information about the visited websites \cite{vilella2020news}.

% SWC
The studies, cited before, mainly focus on the spatial distance between the home and work locations, as it is hard to estimate the travel time, using sporadic \acrshort{CDR} data, though it has seasonal nature due to the human biorhythm.
Moreover, human mobility is highly regular \cite{gonzalez2008understanding,song2010limits}, and the individual activity has a bursty characteristic \cite{barabasi2005origin}.
Jo et al. found that removing the circadian and weekly seasonality, the bursty nature of the human activity remains \cite{jo2012circadian}.

In the digital era, the human sleep wake cycle (\acrshort{SWC}) is also studied using the info-communication systems, such as smartphones \cite{cuttone2017sensiblesleep,aledavood2018social}, websites \cite{yasseri2012circadian,yasseri2013temporal,vilella2020news}, social media \cite{dzogang2017circadian} and call detail records (\acrshort{CDR}) \cite{ahas2010daily,jo2012circadian,aledavood2015daily,lotero2016rich,monsivais2017seasonal,monsivais2017tracking,alakorkko2020circadian}.
Cuttone et al. used screen-on events of the smartphones to study the daily sleep periods \cite{cuttone2017sensiblesleep}, and Aledavood et al. examined the social network of different chronotypes, using the same data set \cite{aledavood2018social}.
Monsivais et al. identified yearly and seasonal patterns in calling activity and resting periods \cite{monsivais2017seasonal,monsivais2017tracking}.
Lotero et al. found connection between temporal patterns and the socioeconomic status of the subscribers, namely the wealthier wakes up later \cite{lotero2016rich}.
Diao et al. found difference in the daily activity between different districts of Boston \cite{diao2016inferring}.

Economic models distinguish city parts such as residential areas, industrial areas, business districts and so on, but that is a rather static, slowly evolving layer of a city. Mobile phone network data has a potential to describe the city structure, via the inhabitants' mobility patterns.
This study focuses on the effect of the \acrshort{SWC} to the city structure. In this regard, it is a continuation of our previous work \cite{pinter2021evaluating}, but this time, the city structure is analyzed by the circadian rhythm of the people who live and work in a given area of the Budapest.

Is it possible to cluster the areas of a city by time, when the activity of the inhabitants, the workers, or the passers-by start their activity in the morning or halt in the evening?
Do city parts have ``chronotypes''? Can neighborhoods or districts be described by the terms ``morningness'' or ``eveningness''? Is there a structural or socioeconomic connection between the areas with the same ``chronotype''?

The goal of this study is to answer these questions, and the contributions of this paper are briefly summarized as follows:
\begin{enumerate}
  \item Introducing wake-up time as an indicator to describe the behavior of a group of subscribers.
  \item Using this indicator to distinguish the areas of Budapest.
  \item Estimating the day length and the working hour length, using the same method.
  \item Demonstrating connection between the wake-up time and the mobility customs.
  \item Identifying correlation between the wake-up time and the socioeconomic status.
\end{enumerate}

The rest of this paper is organized as follows. The utilized data are described in Section~\ref{sec:materials}; then, in Section~\ref{sec:methodology}, the applied methodology is introduced and in Section~\ref{sec:results_and_discussion} the results of this study are presented and discussed. Finally, in Section~\ref{sec:conclusions}, the findings of the paper are summarized, and a conclusion is drawn.

\section{Materials}
\label{sec:materials}

Vodafone Hungary, one of the three mobile phone operators providing services in Hungary, provided two anonymized \acrshort{CDR} data sets for this study. The observation area was Budapest, the capital of Hungary and its agglomeration.
Both observation periods are one month long, the first is June 2016, the second is April 2017.
The nationwide market share of Vodafone Hungary was 25.3\% in 2016 Q2, and 25.5\% in 2017 Q2 \cite{nmhh_mobile_market_report}.

The communication between a cellular device and the mobile phone network can be divided into two categories: (i) An administrative communication maintaining the connection with the service, for example, registration of the cell-switching that can be called passive communication. (ii) When the device actively uses the network for voice calls, message or data transfer, that can be called active communication.
The available data contains only the active communication, which is sparser, so it cannot be used to track continuous movements.

The obtained data contains a hashed value to identify the \acrshort{SIM} (Subscriber Identity Module), a timestamp that is truncated to 10 s, and an ID of the cell. Thus, a subscriber can be mapped to a geographic location in a given time. These are extended with the type of the customer (business, consumer), the type of the subscription (prepaid, postpaid), the age and gender of the subscriber and the Type Allocation Code (\acrshort{TAC}) of the device. The \acrshort{TAC} is the first eight digits of the \acrshort{IMEI} (International Mobile Equipment Identity) number, and allocated by the GSM Association and uniquely identifies the mobile phone model.
These values are also present in every record, so for example the device changes can be tracked as well.

Both the type of the activity (voice call, message, data transfer) and the direction (incoming, outgoing) were omitted from the data, and there was no data provided by the operator to resolve the \acrshort{TAC}s to manufacturer and model.

The data was processed using the same framework as in out previous works \cite{pinter2021evaluating,pinter2021analyzing}.
During the data cleaning process, the wide format was normalized. The \acrshort{CDR} table contained only the \acrshort{SIM} ID, the timestamp, and the cell ID. For the subscriber and subscription related information,
a separate table was formed, and another table was created to track the device changes of the subscriber.

Another table has been created for the cell information, including a cell ID, the geographic location of the cell centroid and base station. The cell centroid is an estimation based on a momentary state, especially with the UMTS cells, that apply a load balancing mechanism, called ``cell breathing'', which can change the geographic size of the serving area. The heavily loaded cells shrink, and the neighboring ones grow to compensate \cite{al2003performance}.

\subsection{June 2016}
\label{sec:vod201606}

\begin{figure}
    \includegraphics[width=\linewidth]{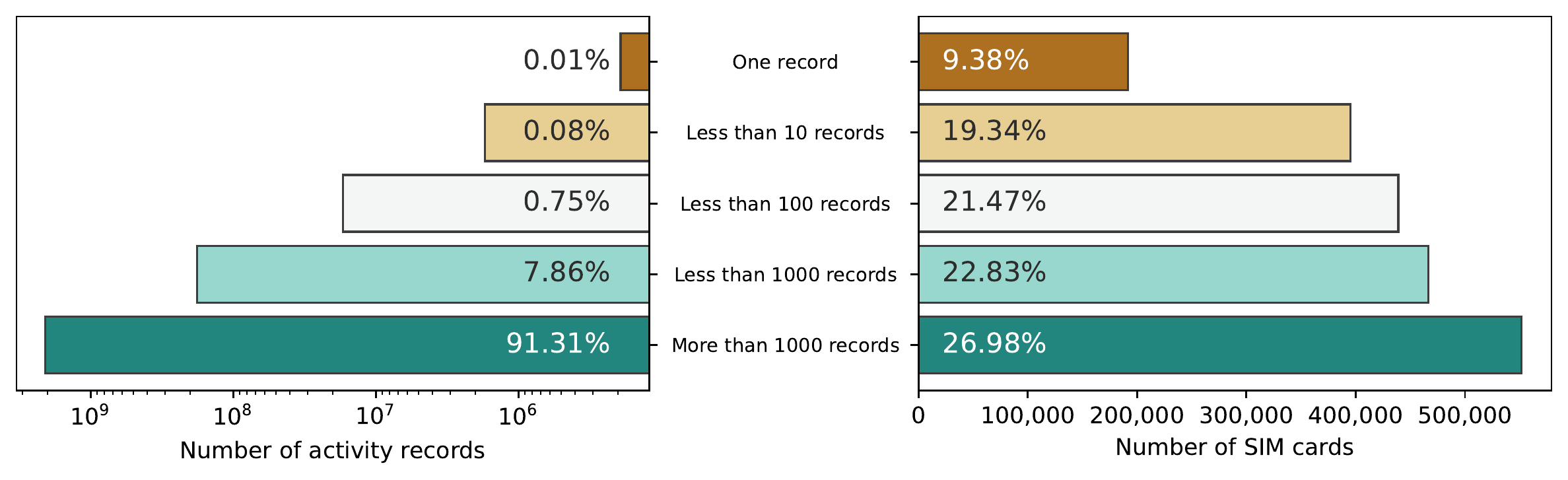}
    \caption{The \acrshort{SIM} cards, in the ``June 2016'' data set, categorized by the number of activity records.
    The \acrshort{SIM} cards with more than 1000 activity records (26.98\% of the \acrshort{SIM} cards) provide the majority (91.31\%) of the activity.}
    \label{fig:vod201606_sim_activity}
\end{figure}

This data set includes \num{2291246932} records from \num{2063005} unique \acrshort{SIM} cards, during the observation period of June 2016.
Figure~\ref{fig:vod201606_sim_activity} shows the activity distribution between the activity categories of the \acrshort{SIM} cards. The dominance of the last category, \acrshort{SIM} cards with more than 1000 activity records, is even more significant. It can be seen that almost 27\% of the \acrshort{SIM} cards produce more than 91\% of the total activity.

\begin{figure}
    \includegraphics[width=\linewidth]{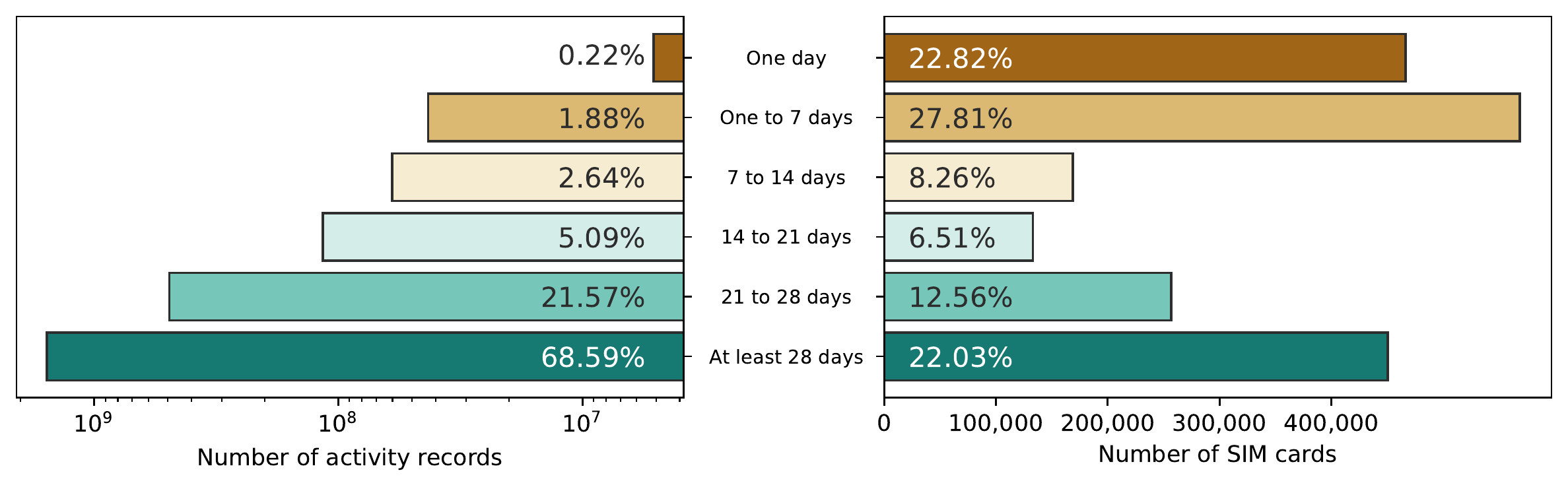}
    \caption{\acrshort{SIM} card distribution of the ``June 2016'' data set, by the number of active days.}
    \label{fig:vod201606_activity_by_days}
\end{figure}

Figure~\ref{fig:vod201606_activity_by_days} shows the \acrshort{SIM} card distribution according to the number of active days. Only 34.59\% of the \acrshort{SIM} cards had activity on at least 21 different days.
In total, \num{241824} \acrshort{SIM} cards (11.72\%) appeared on at least two days, but~the difference between the first and the last activity was not more than seven days. This may indicate the presence of tourists. High levels of tourism are usual during this part of the year.

Type Allocation Codes are provided for every record because a subscriber can change their device at any time. Naturally, most of the subscribers (\num{95.71}\%) used only one device during the whole observation period.

While the subscription details were available for every \acrshort{SIM} card, the subscriber information was missing in slightly more than 40\% of the cases, presumably because of the subscribers' preferences regarding the use of their personal data.

Although the data contained cell IDs, only the base station locations, where the cell antennas were located, were known.
As a base station usually serve multiple cells, these cells were merged by the serving base stations. After~the merge, 665 locations (sites) remained with known geographic locations. To estimate the area covered by these sites, Voronoi Tessellation was performed on the locations. This is a common practice \cite{csaji2013exploring,pappalardo2016analytical,vanhoof2018comparing,zagatti2018trip,novovic2020uncovering} in \acrshort{CDR} processing.

\subsection{April 2017}
\label{sec:vod201704}

\begin{figure}
    \includegraphics[width=\linewidth]{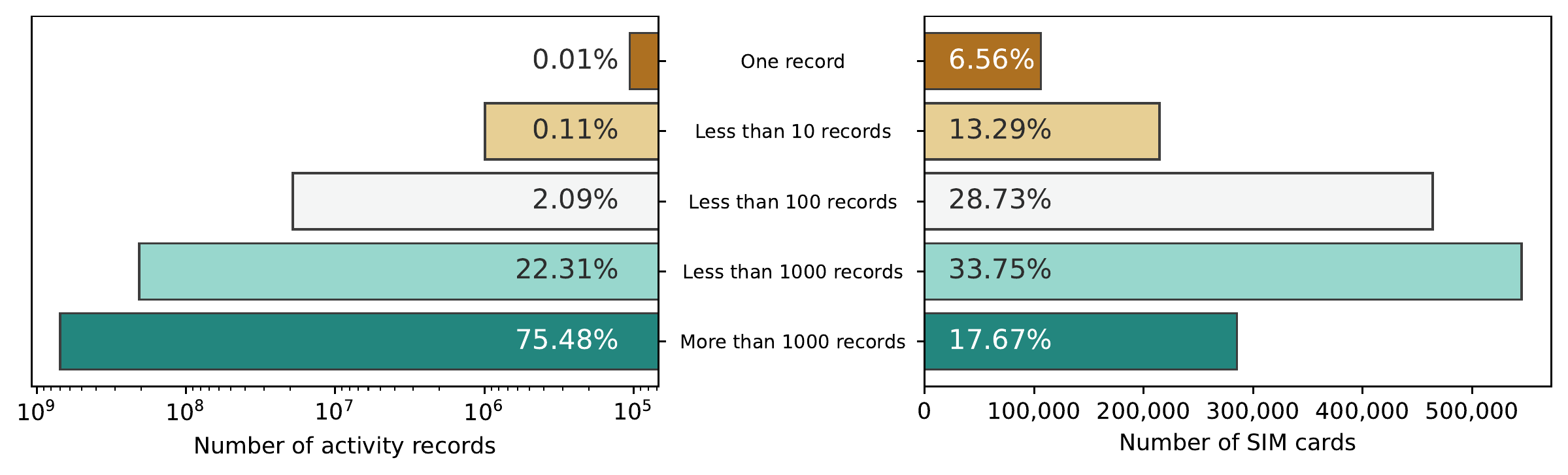}
    \caption{The \acrshort{SIM} cards in the ``April 2017'' data set categorized by the number of activity records. The \acrshort{SIM} cards over a thousand records (17.7\%) provide the majority (75.48\%) of the activity.}
    \label{fig:vod201704_activity_categories}
\end{figure}

\begin{figure}
    \includegraphics[width=\linewidth]{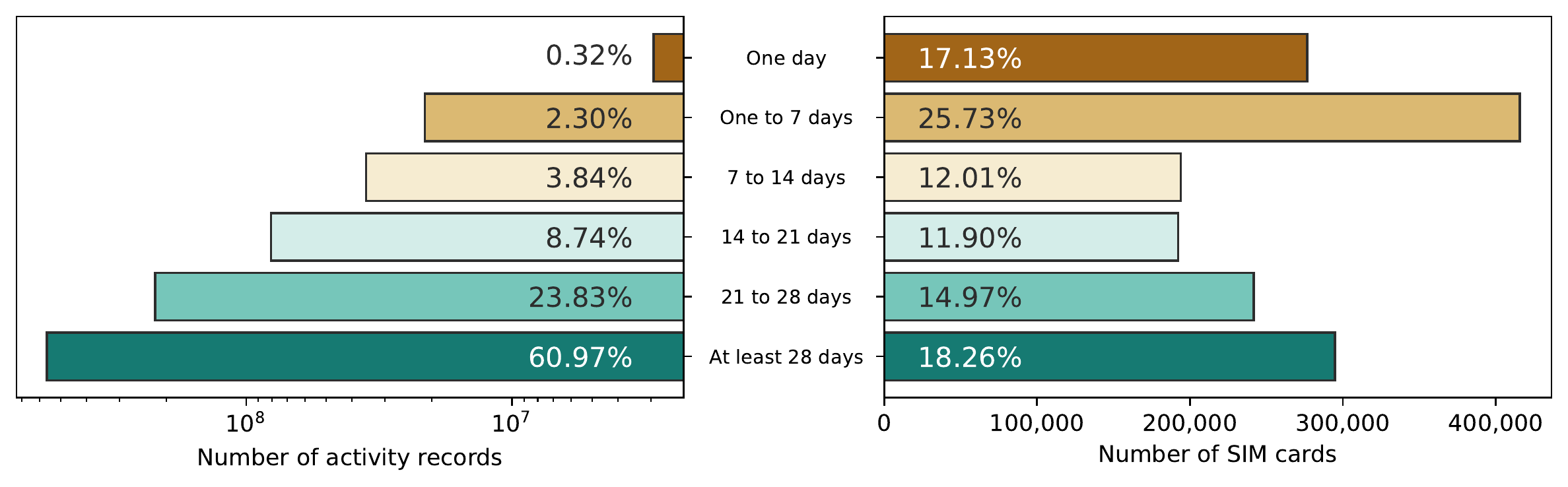}
    \caption{\acrshort{SIM} card distribution of the ``April 2017'' data set, by the number of active days.}
    \label{fig:vod201704_activity_by_days}
\end{figure}

\begin{figure}
    \begin{subfigure}[t]{\linewidth}
        \includegraphics[width=\linewidth]{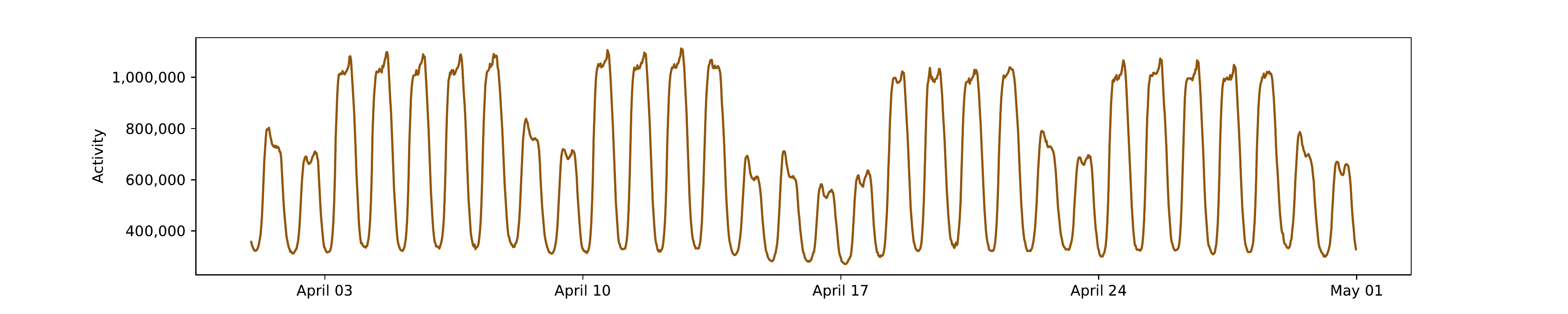}
        \captionsetup{position=bottom,justification=centering}
        \caption{}
        \label{fig:timeseries}
    \end{subfigure}
    \vfill
    \vspace{0.5em}
    \begin{subfigure}[t]{\linewidth}
        \includegraphics[width=\linewidth]{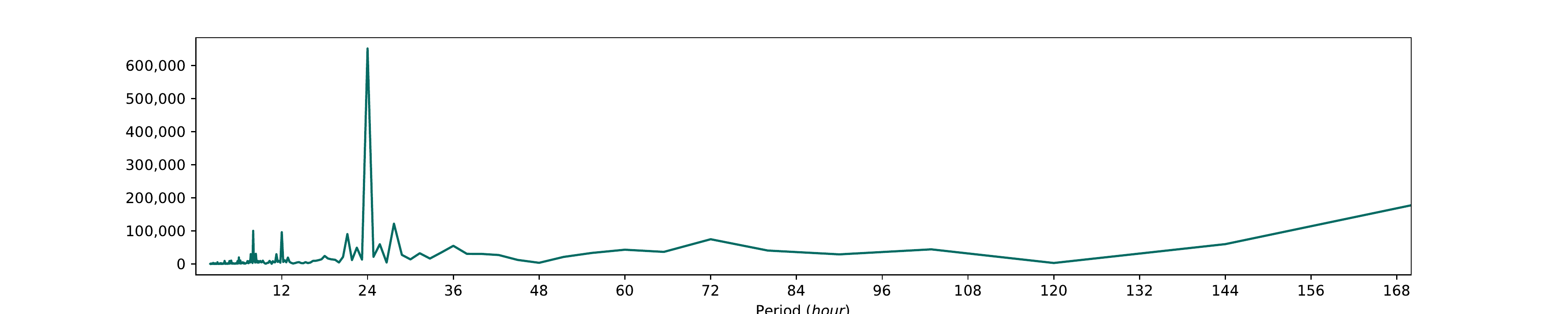}
        \captionsetup{position=bottom,justification=centering}
        \caption{}
        \label{fig:fourier}
    \end{subfigure}
    \caption{The mobile phone network activity (\textbf{a}) during the observation period of April 2017, and its Fourier decomposition (\textbf{b}).}
    \label{fig:vod201704_seasonality}
\end{figure}

The second \acrshort{CDR} data set contains \num{955035169} activity records, from \num{1629275} \acrshort{SIM} cards, and the observation period is April 2017.
Figure~\ref{fig:vod201704_activity_categories}, shows the activity distribution between the activity categories of the \acrshort{SIM} cards. Only 17.67\% of all the \acrshort{SIM} cards, that have more than 1000 activity records, provide the majority (75.48\%) of the mobile phone activity during the observation period.
Figure~\ref{fig:vod201704_activity_by_days}, shows the distribution of the \acrshort{SIM} cards by the number of active days. Only about one-third (33.23\%) of the \acrshort{SIM} cards have activity on at least 21 different days.
In spite of the relatively large number of \acrshort{SIM} cards, present in the data, most of them are not active enough to provide enough information about their mobility habits.
Figure~\ref{fig:vod201704_seasonality}a, shows the mobile phone network activity distribution during the observation period, and Figure~\ref{fig:vod201704_seasonality}b its Fourier decomposition to highlight the seasonality of the data. As expected, this 30-day dataset has a 24-hour periodicity.
In this study, mainly this data set was used, the ``June 2016'' is only used as a reference for some analyses (e.g., in Section~\ref{sec:day_length}).

\subsection{Data for the Socioeconomic Indicators}
\label{sec:ses_data}

\begin{figure*}
    \begin{subfigure}[t]{0.33\linewidth}
        \includegraphics[width=\linewidth]{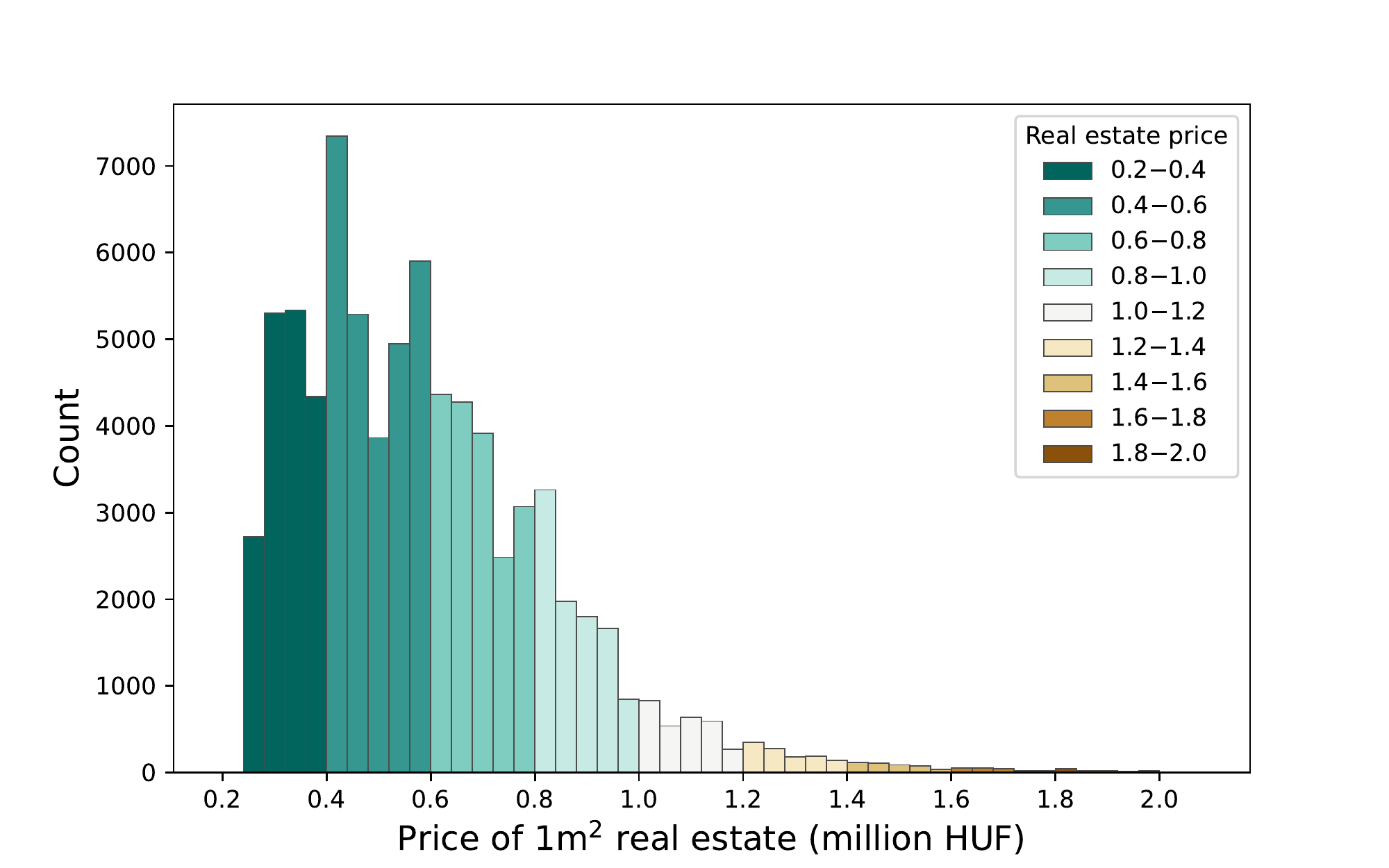}
        \captionsetup{position=bottom,justification=centering}
        \caption{}
        \label{fig:ingatlancom_histogram}
    \end{subfigure}
    \hfill
    \begin{subfigure}[t]{0.33\linewidth}
        \includegraphics[width=\linewidth]{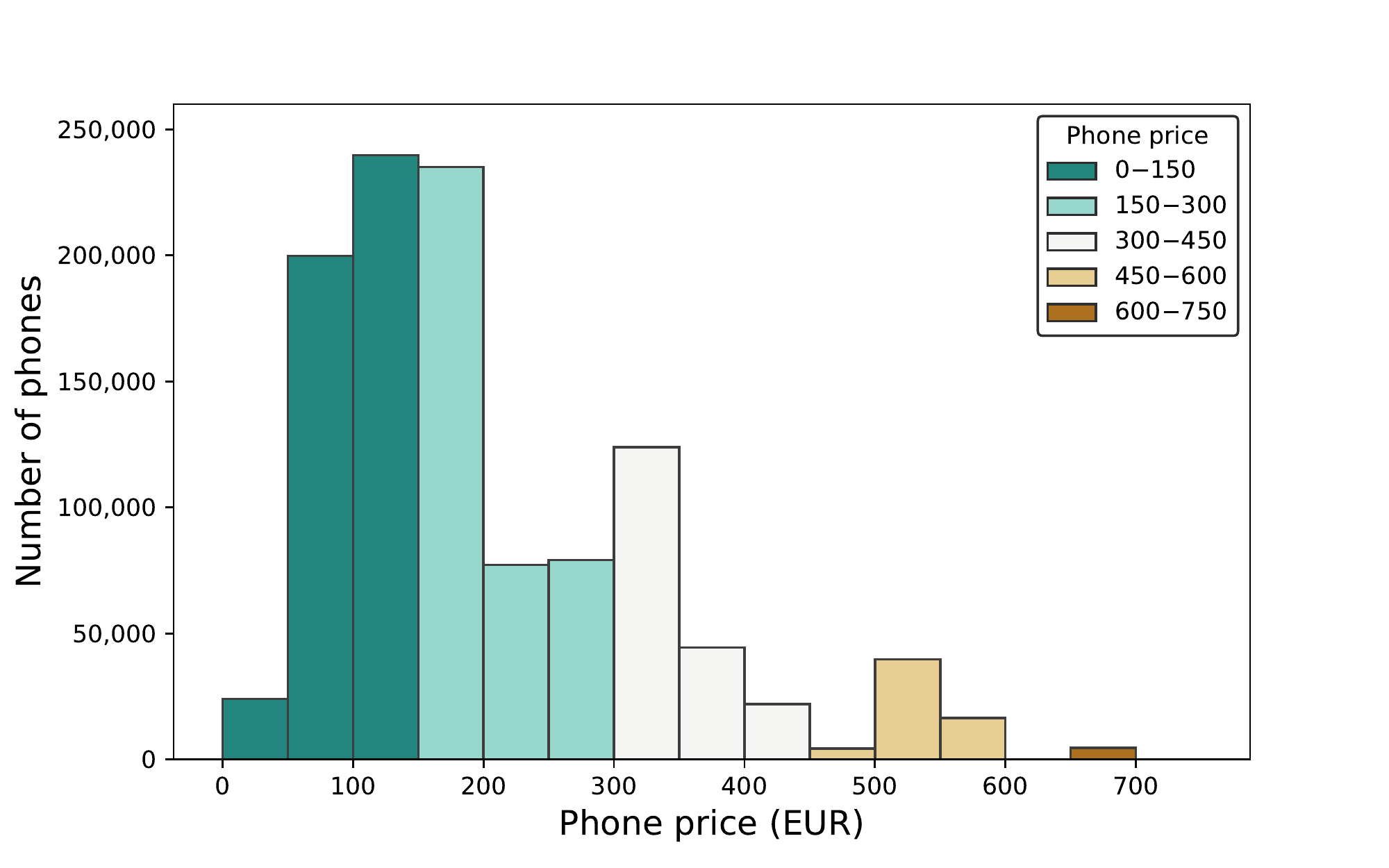}
        \captionsetup{position=bottom,justification=centering}
        \caption{}
        \label{fig:pp_hist}
    \end{subfigure}
    \hfill
    \begin{subfigure}[t]{0.33\linewidth}
        \includegraphics[width=\linewidth]{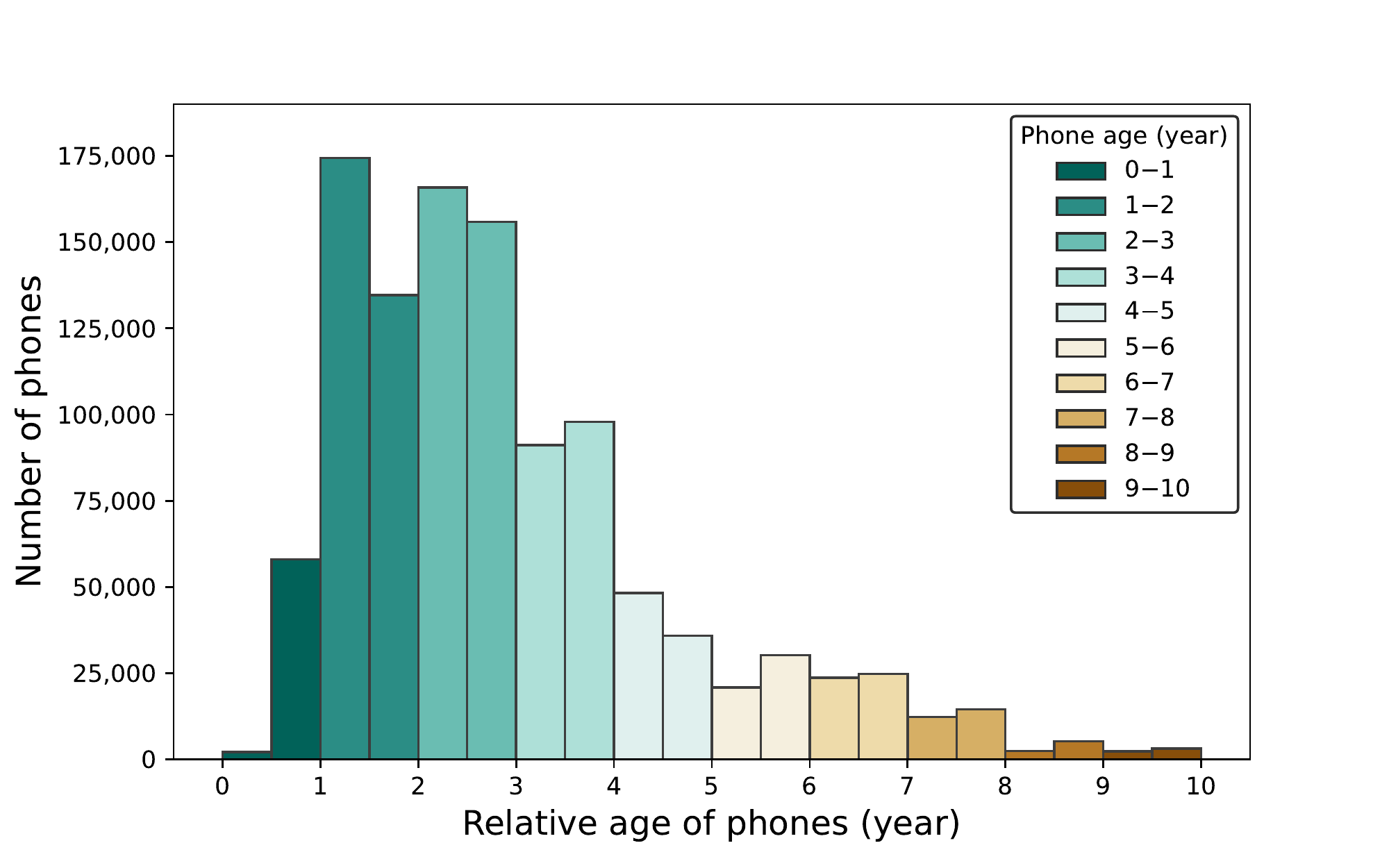}
        \captionsetup{position=bottom,justification=centering}
        \caption{}
        \label{fig:pa_hist}
    \end{subfigure}
    \caption{Distribution of the real property prices (\textbf{a}), the mobile phone prices (\textbf{b}), and the mobile phone relative ages (\textbf{c}).}
    \label{fig:property_and_phone_price}
\end{figure*}

In a previous work \cite{pinter2021evaluating}, real estate prices a were used to characterize the socioeconomic status. The \textit{ingatlan.com} estate selling website provided more than 60 thousand estate locations with floor spaces and selling prices from Budapest and Pest county. These prices were used to describe the property prices at the subscribers' home locations.
Figure~\ref{fig:ingatlancom_histogram}, shows the distribution of the normalized housing prices from the advertisements.

Our former paper \cite{pinter2021analyzing} was focusing on the price and the relative age of the subscribers' mobile phone were utilized as a socioeconomic indicator.
The data, used for resolving the \acrshort{TAC}s to hardware vendor and model, was provided by \textit{51Degrees}.
With the model and type of the device, in which the \acrshort{SIM} cards operate, can also be used to remove those \acrshort{SIM} cards, that is not used in mobile phones, thus do not represent a human.
The cell phone prices and the release dates were obtained from \textit{GSMArena} \cite{mohit_gsmarena}.
The relative age of a phone is determined as the difference of the date of the \acrshort{CDR} data set (e.g., April 2017) and the release date of the phone.
Figure~\ref{fig:pa_hist}, shows that most of the devices are one to three years old, but there are some really outdated (over eight years) phones in use.
Figure~\ref{fig:pp_hist}, displays the distribution of the cell phone prices.
Although, the data source is supposed to contain launch prices, the validation analyzes (Appendix~\ref{app:iphone}) implies that the phone prices are depreciated.

\section{Methodology}
\label{sec:methodology}

In In a previous work \cite{pinter2021evaluating}, a framework was introduced to process the mobile phone network data, described in Section~\ref{sec:materials}.
The~\acrshort{CDR}s were normalized, cleaned and the mobility metrics (Section \ref{sec:mobility_metrics}) were determined for every subscriber.

\subsection{Home and Work Locations}
\label{sec:home_work}

Most of the inhabitants in cities spend significant time of a day at two locations: their homes and work places. In order to find the relationship between these most important locations, first their positions of these locations have to be determined.
There are a few approaches used to find home locations via mobile phone data analysis  \cite{ahas2010using,xu2015understanding,vanhoof2018assessing,mamei2019evaluating,pappalardo2021evaluation}.

Our solution is similar to the most common approach.
The most frequent cell where a \acrshort{SIM} card is present during working hours is considered as the work location, on workdays between 09:00 and 16:00.
The home location is calculated as the most frequent cell where a \acrshort{SIM} card is present during the evening and the night on workdays (from 22:00 to 06:00) and all day on holidays. Although people do not always stay at home on the weekends, it is assumed that a significant amount of activity is generated from their home locations.
In \cite{pinter2021evaluating}, we used census data to indirectly, via commuting trends, validate the estimated home and work locations.

\subsection{Mobility Metrics}
\label{sec:mobility_metrics}

Along with the Home and Work locations, the Radius of Gyration and the Entropy are commonly used \cite{pappalardo2015returners,pappalardo2016analytical,xu2018human,cottineau2019mobile,mamei2019evaluating,bhattacharya2019social,pinter2021evaluating,lucchini2021living} indicators of human mobility, and were also determined for every subscriber.

The Radius of Gyration~\cite{gonzalez2008understanding} is the radius of a circle in which an individual (represented by a \acrshort{SIM} card) can usually be found.
It was originally defined in Equation~(\ref{eq:gyration}), where $L$ is the set of locations visited by the individual, $r_{cm}$ is the center of mass of these locations, and $n_i$ is the number of visits to the i-th location.

\begin{equation}
    \label{eq:gyration}
    r_g = \sqrt{\frac{1}{N} \sum_{i \in L}{n_i (r_i - r_{cm})^2}}
\end{equation}

The entropy characterizes the diversity of locations visited using an individual's movements, defined as Equation~(\ref{eq:entropy}), where $L$ is the set of locations visited by an individual, $l$ represents a single location, $p(l)$ is the probability of an individual being active at a location $l$, and $N$ is the total number of activities taken part in by an individual \cite{pappalardo2016analytical,cottineau2019mobile}.

\begin{equation}
    \label{eq:entropy}
    e = - \frac{\sum_{l \in L}{p(l) \log p}}{\log N}
\end{equation}

\subsection{Wake-up Time}
\label{sec:wakeup_time}

\begin{figure}
  \centering
  \includegraphics[width=.75\linewidth]{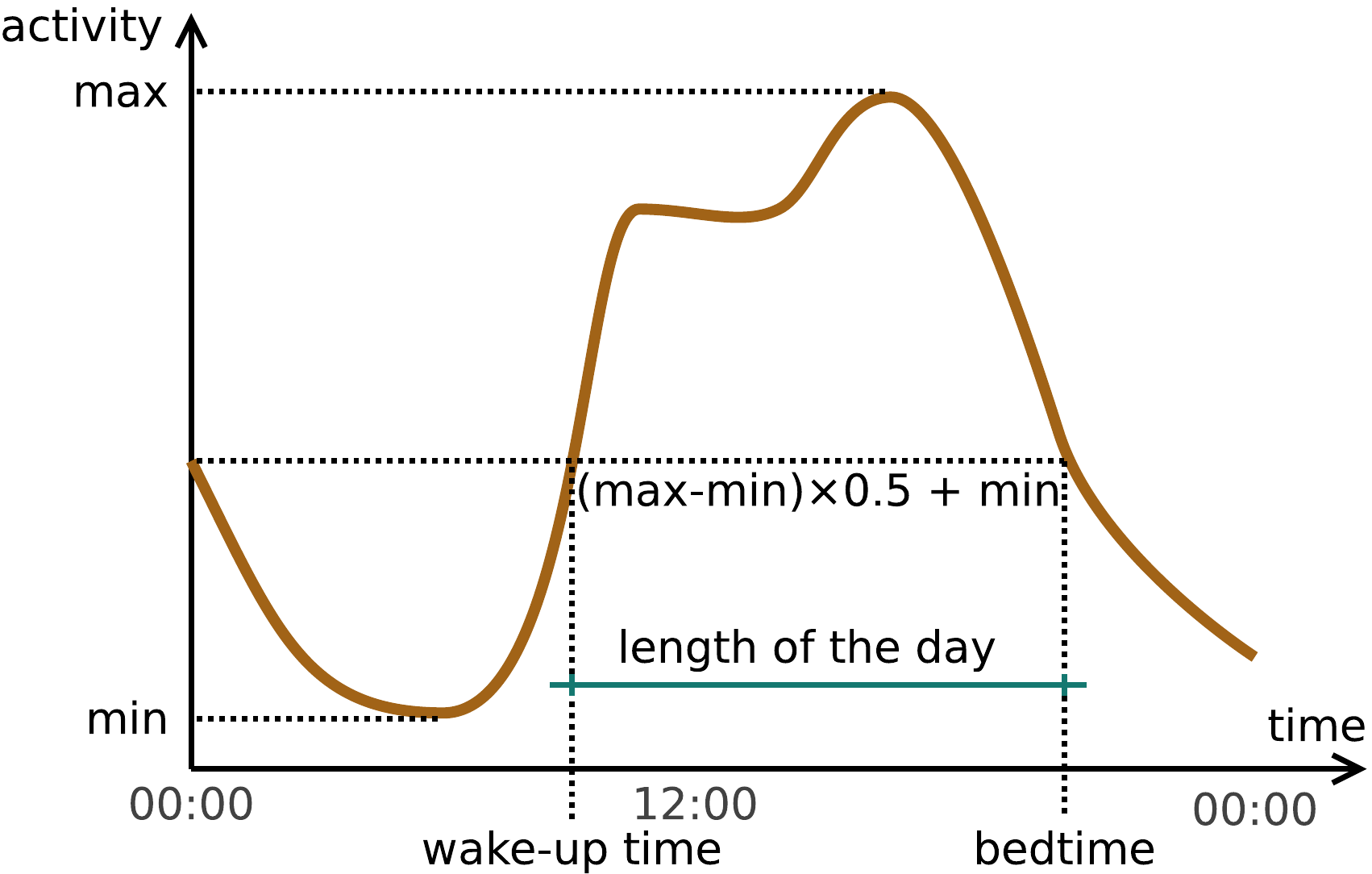}
  \caption{Calculation of the wake-up and the bedtimes}
  \label{fig:wakeup_calc}
\end{figure}

The Call Detail Records are aggregated for every cell and 10 minute time intervals. Then, moving average is applied with the window of 12.
The minimum of the aggregated records is usually in the middle of the night, and the maximum is in the afternoon. To determine the wake-up time, the positive edge of the curve needs to be detected, when the number of the mobile phone activity increases drastically in a short period of time.
The time, that is considered as the wake-up time, is simply where the activity value reaches the mean of the minimum and the maximum of the curve. Figure~\ref{fig:wakeup_calc}, illustrates the concept.
Equation~\ref{eq:wakeup}, defines the mean of the activity edge, where $a_{min}$ and $a_{max}$ is the minimum and the maximum of the (daily) activity level, respectively.

\begin{equation}
    \label{eq:wakeup}
    m = (a_{max}-a_{min})*0.5 + a_{min}
\end{equation}

This process is repeated for every day, then the median of the daily wake-up times are determined.
Analogously to the wake-up time, the bedtime can be calculated to describe when the mobile phone activity decreases significantly, as selecting the mean value on the negative edge.
It has to be noted, that these values are naturally not the actual times when people wake up or fall asleep. Those moments cannot be determined using simply the mobile phone network.
Using the screen-on events of the phone \cite{cuttone2017sensiblesleep} can be much closer to the actual values. Especially in the case of the wake-up time, if the phone is used as an alarm clock.
In spite of this, it is supposed that this approximating method can reveal the rough tendencies of the daily routine. Still, the terms ``wake-up time'' and ``bedtime'' are used to refer to the time of the positive/rising and negative/falling edges of the daily activity curve, respectively.

\subsection{Aggregation of the Subscribers}
\label{sec:grouping}

As the type of the phone activity is unknown, the wake-up time of the individual is hard to estimate. Since making a call requires to be awake (whereas message could be received and data could be transferred autonomously), knowing which activities represent phone calls would provide more accurate information about when people are certainly awake. Instead of this, the significant activity increase is to be used to identify the wake-up time.

Furthermore, activity of a single device is also sporadic and does not provide enough data to identify an activity increase for every day.
To mitigate the downside of the sporadic nature of the data, the devices are grouped, that can be performed in two ways: (i) calculate the wake-up time of an area (a cell or a group of cells) or (ii) the inhabitants of an area.

\begin{figure}[th]
    \begin{subfigure}[t]{0.48\linewidth}
        \includegraphics[width=\linewidth]{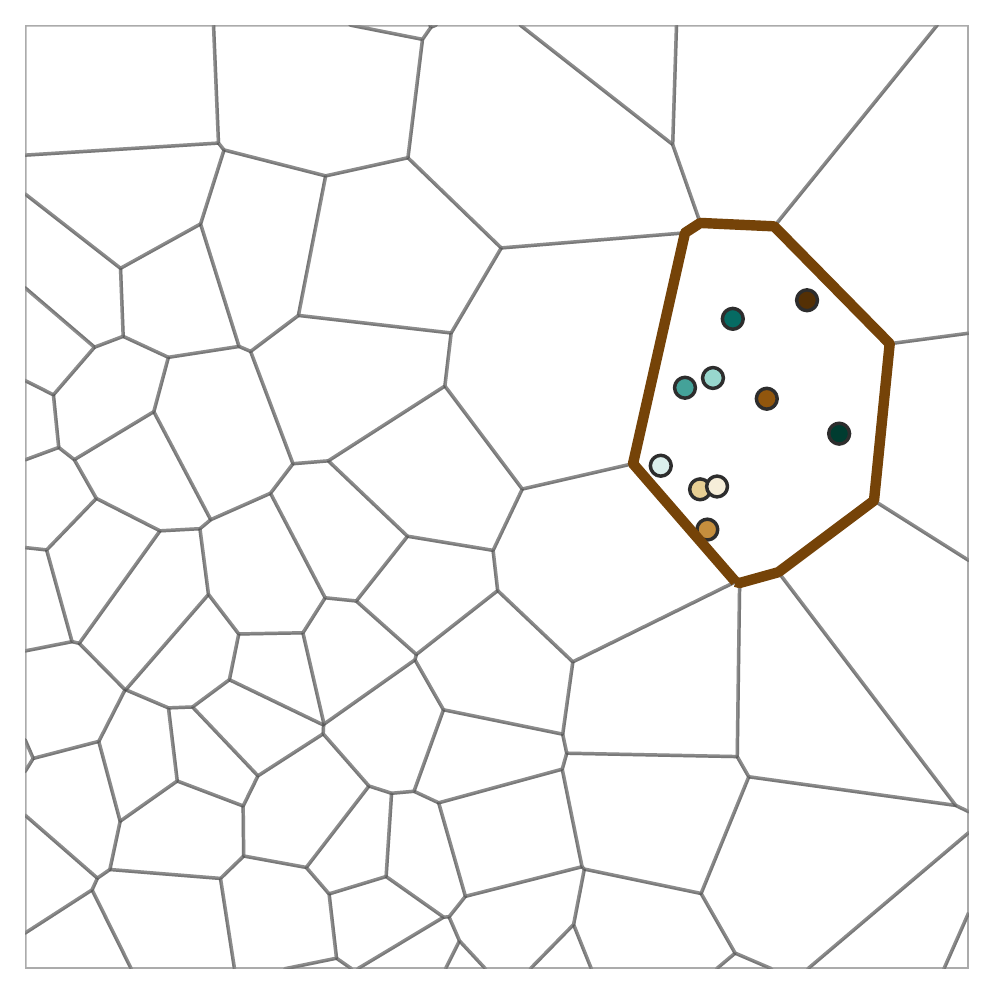}
        \captionsetup{position=bottom,justification=centering}
        \caption{}
        \label{fig:cell_based_grouping}
    \end{subfigure}
    \hfill
    \begin{subfigure}[t]{0.48\linewidth}
        \includegraphics[width=\linewidth]{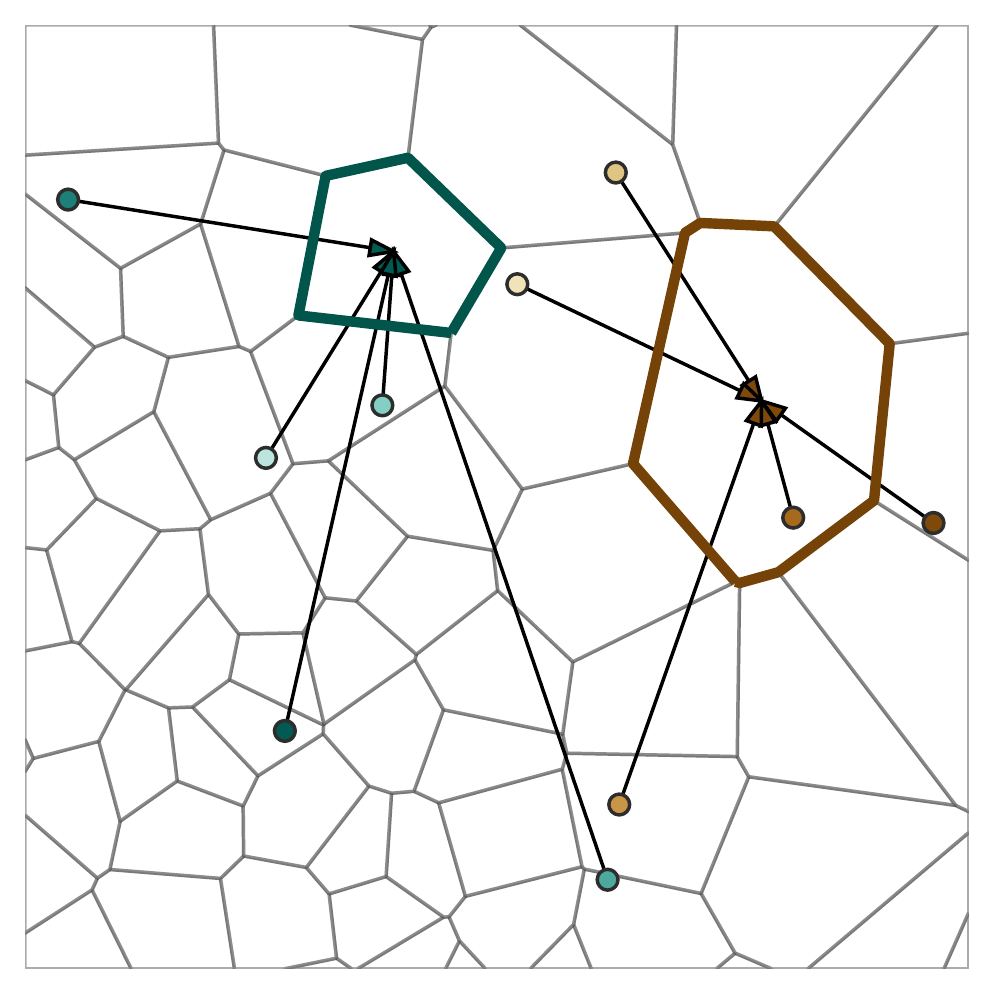}
        \captionsetup{position=bottom,justification=centering}
        \caption{}
        \label{fig:dweller_based_grouping}
    \end{subfigure}
    \caption{Visualizing the difference between the cell-based (\textbf{a}) and the inhabitant-based (\textbf{b}) approaches. The former considers subscribers present in a given cell wherever they live. The latter aggregates activity of the inhabitants to the home cell, regardless where the activity occurred.}
    \label{fig:grouping}
\end{figure}

In the first approach, the activity records are aggregated that takes place in a given cell, regardless which \acrshort{SIM} cards produce them. In the case of the second version, those activity records are used, that produced by the inhabitants of the given cell, regardless where the activity took place.
The first approach could be called cell-based grouping and the latter inhabitant-based grouping. The two approaches are illustrated in Figure~\ref{fig:grouping}.
Cells can be grouped further to examine a larger area (residential, suburb, district, etc.).

\section{Results and Discussion}
\label{sec:results_and_discussion}

In this section, the results have been summarized and discussed, and the limitations and the future work has been included.

\subsection{Inhabitant-based Approach}
\label{sec:dweller_based_wakeup}

\begin{figure*}
    \begin{subfigure}[ht]{0.33\linewidth}
        \includegraphics[width=1\linewidth]{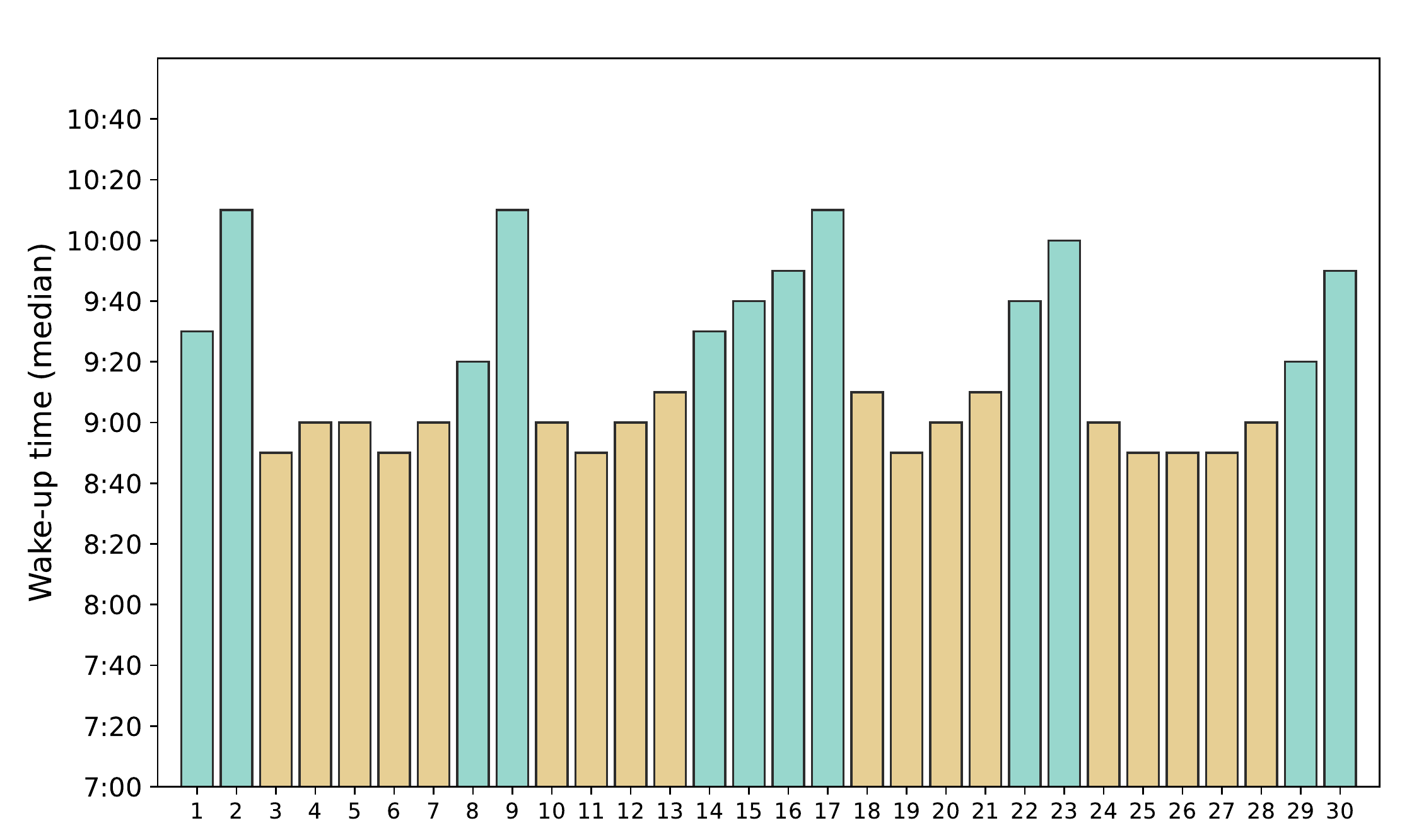}
        \captionsetup{position=bottom,justification=centering}
        \caption{}
        \label{fig:vod201704_daily_wakeups}
    \end{subfigure}
    \hfill
    \begin{subfigure}[ht]{0.33\linewidth}
        \includegraphics[width=1\linewidth]{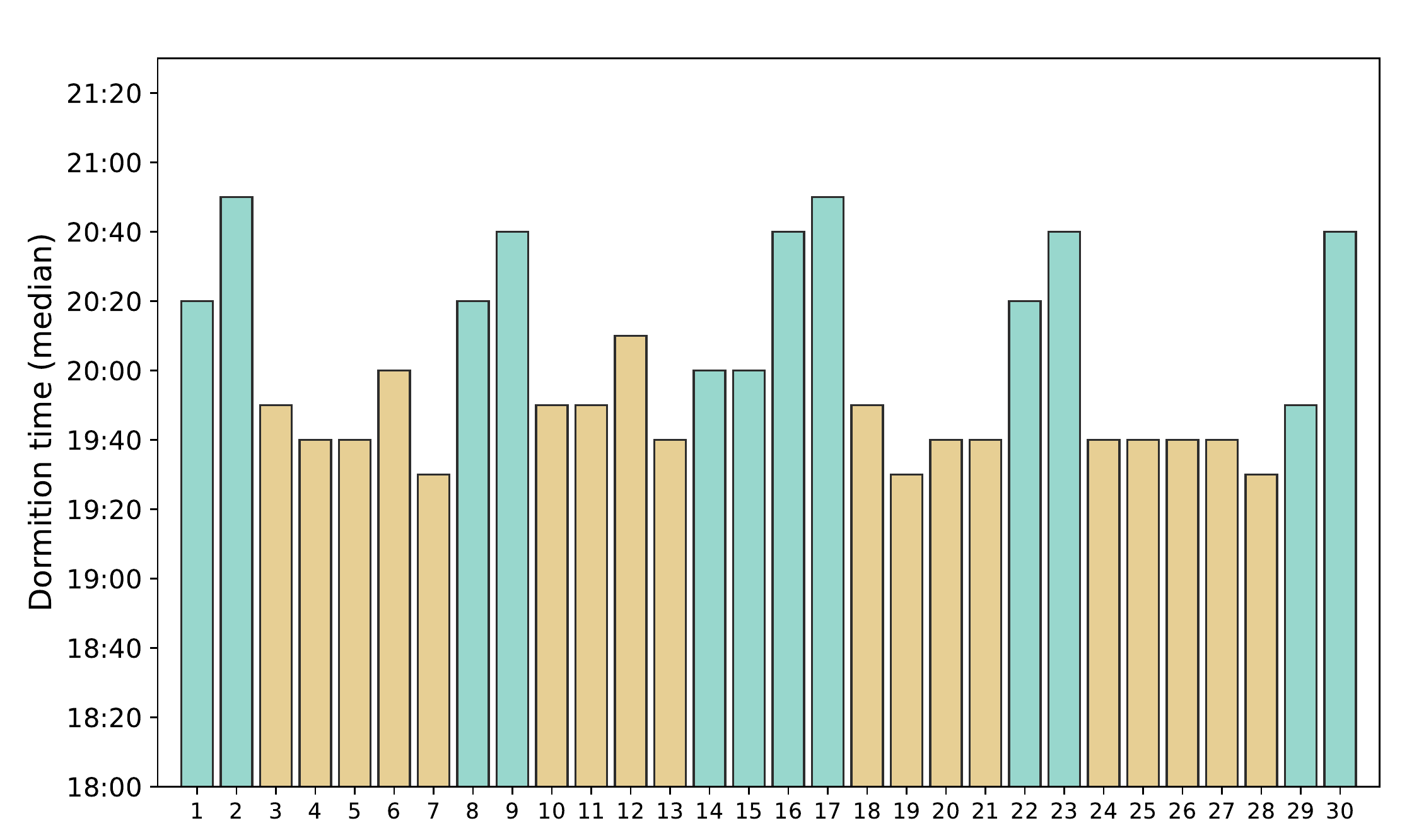}
        \captionsetup{position=bottom,justification=centering}
        \caption{}
        \label{fig:vod201704_daily_dormitions}
    \end{subfigure}
    \hfill
    \begin{subfigure}[ht]{0.33\linewidth}
      \includegraphics[width=1\linewidth]{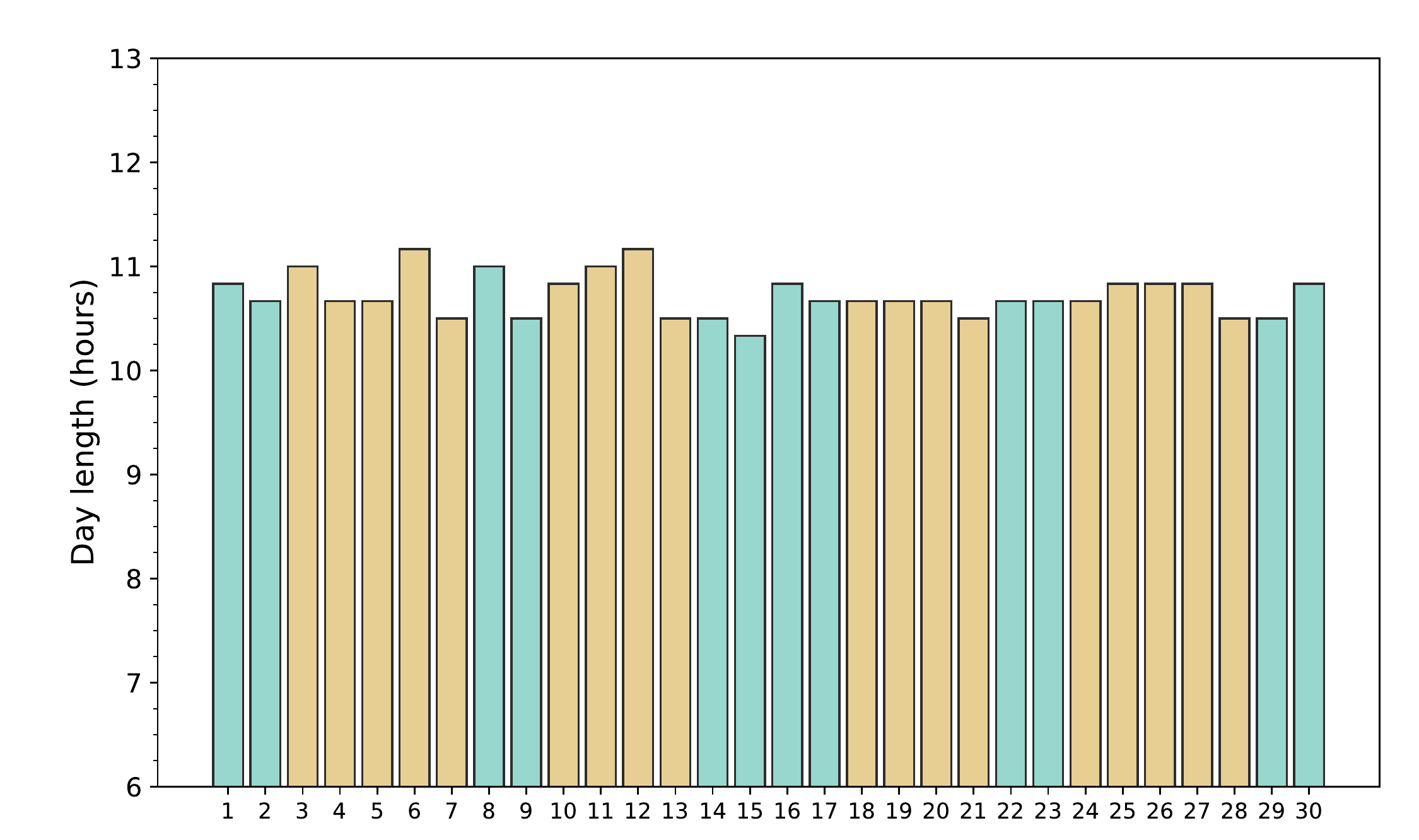}
      \captionsetup{position=bottom,justification=centering}
      \caption{}
      \label{fig:vod201704_day_lengths}
    \end{subfigure}

  \vfill
  \vspace{0.75em}

  \begin{subfigure}[th]{0.33\linewidth}
      \includegraphics[width=1\linewidth]{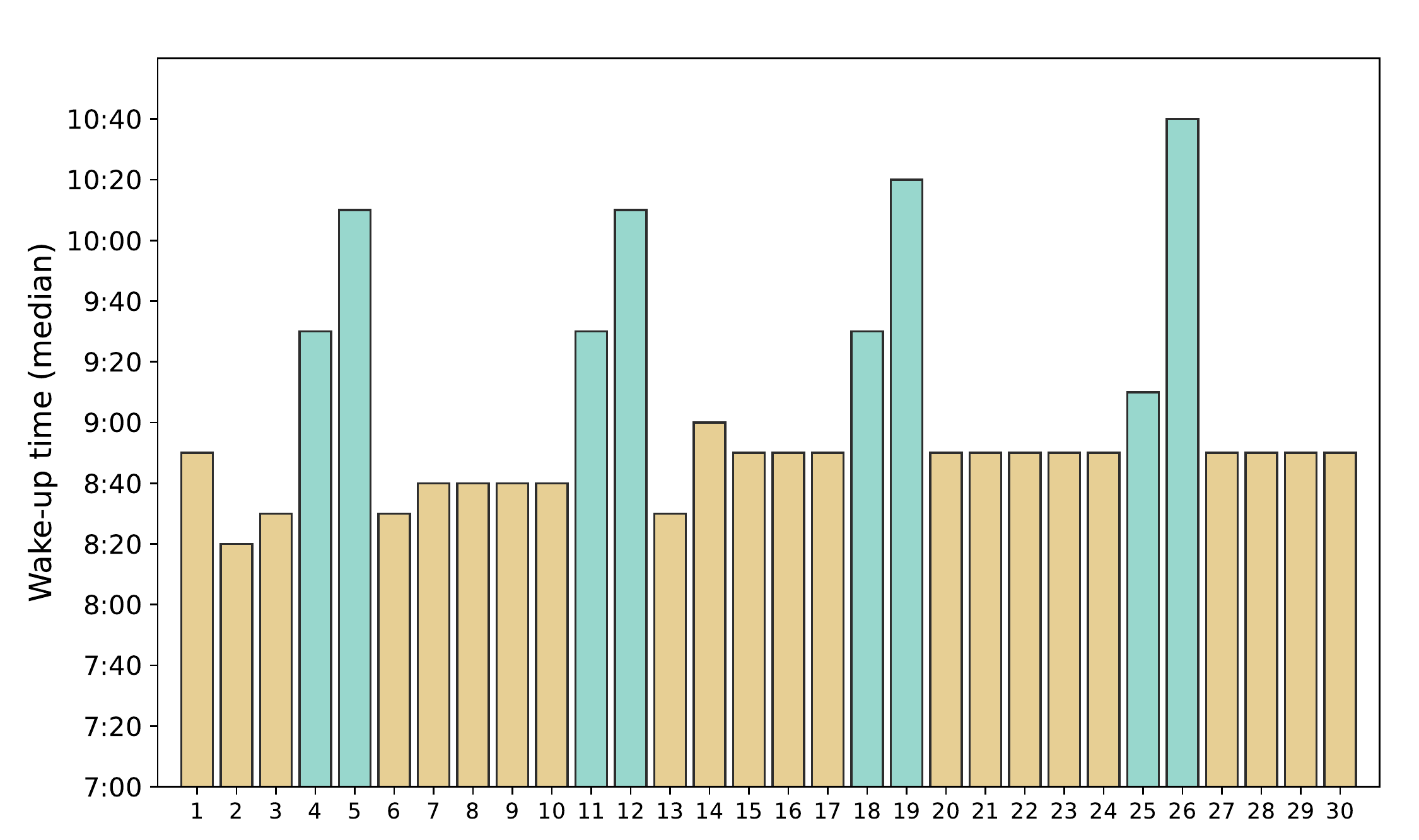}
      \captionsetup{position=bottom,justification=centering}
      \caption{}
      \label{fig:vod201606_daily_wakeups}
  \end{subfigure}
  \hfill
  \begin{subfigure}[th]{0.33\linewidth}
      \includegraphics[width=1\linewidth]{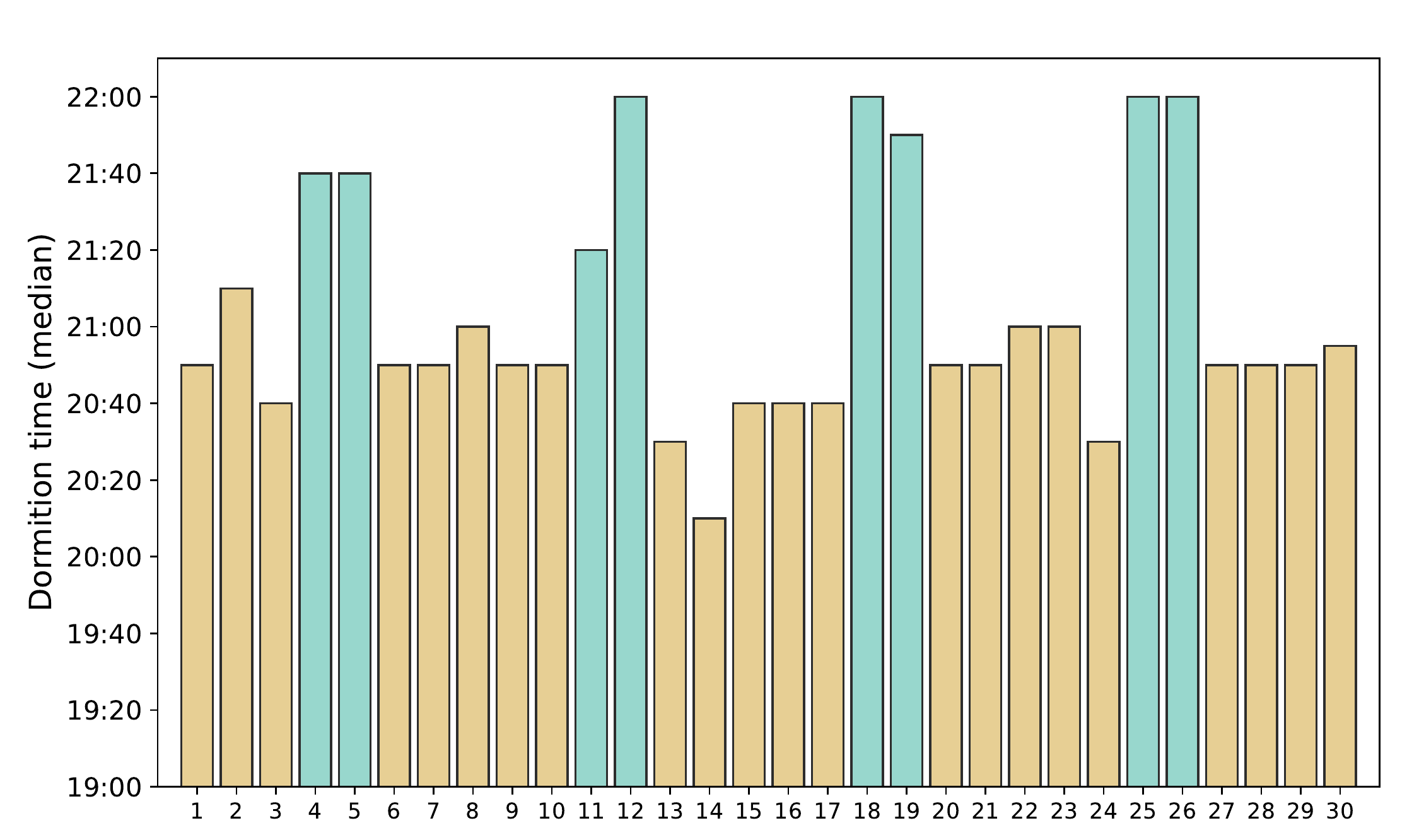}
      \captionsetup{position=bottom,justification=centering}
      \caption{}
      \label{fig:vod201606_daily_dormitions}
  \end{subfigure}
  \hfill
  \begin{subfigure}[th]{0.33\linewidth}
    \includegraphics[width=1\linewidth]{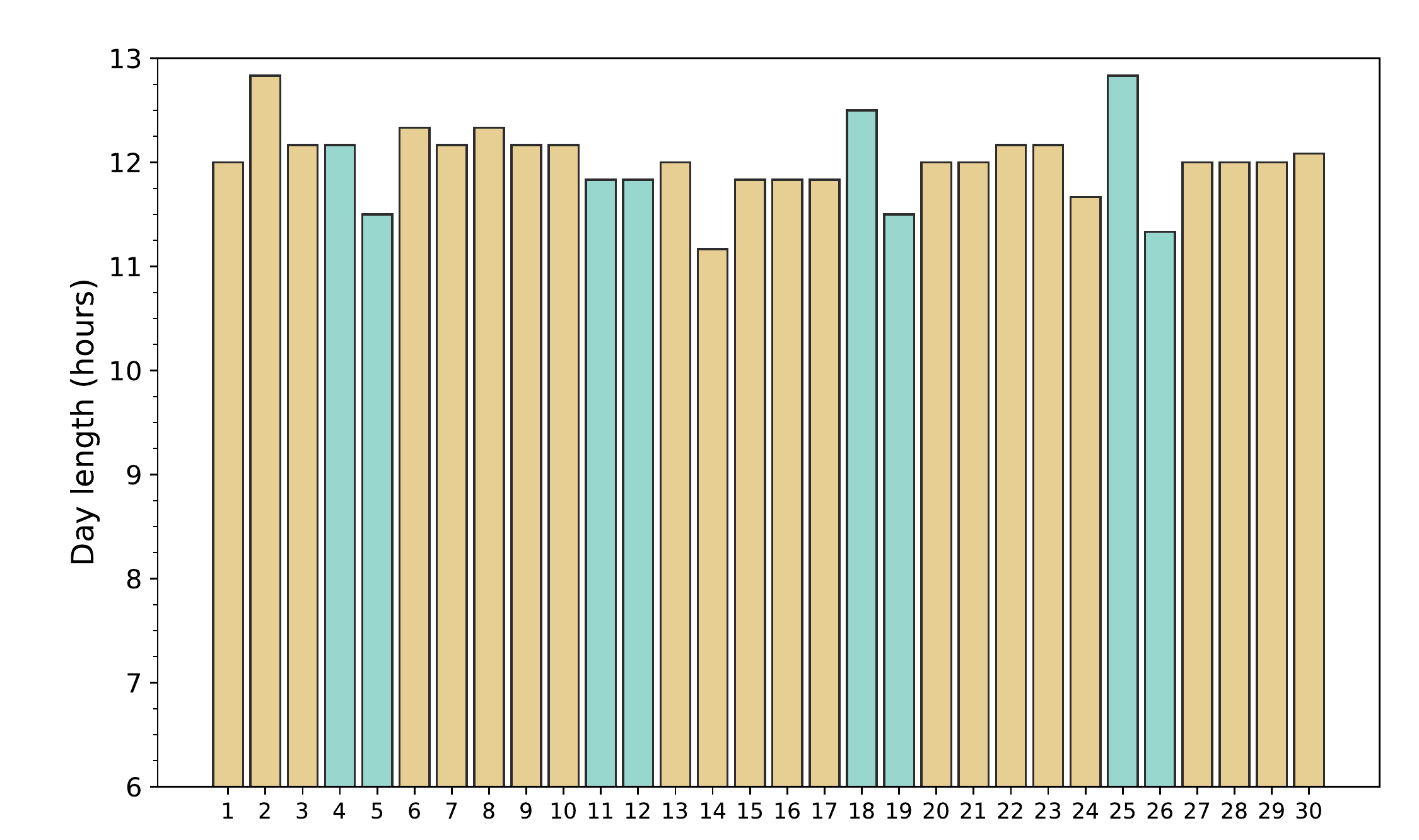}
    \captionsetup{position=bottom,justification=centering}
    \caption{}
    \label{fig:vod201606_day_length}
  \end{subfigure}
  \caption{Daily wake-up times (\textbf{a},\textbf{d}), bedtimes (\textbf{b},\textbf{e}), and the day lengths (\textbf{c},\textbf{e}), for the ``April 2017'' (\textbf{a}--\textbf{c}) and the ``June 2016'' (\textbf{e}--\textbf{e}) dataset. Holidays are represented by a different color.}
  \label{fig:daily_wakeup_dormition_and_daylength}
\end{figure*}

Based on the results of the ``April 2017'' data set (Figure \ref{fig:vod201704_daily_wakeups} and \ref{fig:vod201704_daily_dormitions}), the wake-up time almost always around 7:10 on workdays. On Sundays (or Easter Monday in the case of the long holiday), it is later by about one hour. On Saturdays (and the other days of the holiday), it is later by about 30--40 minutes compared to an average workday.

The bedtimes are not so even, but also have a nice trend. The activity decreases between 19:40 and 20:10, on workdays, but the holiday values shifted by 30--50 minutes, similarly to the wake-up times. Interestingly, as bedtimes follows the wake-up times on the weekends, it results that the average day length remains approximately the same.
Figure~\ref{fig:vod201704_day_lengths}, shows the day lengths, calculated as a difference of the bedtime and the wake-up time, in minutes. The day lengths are between 12 hours 30 minutes and 13 hours 10 minutes. In average, the day length is 12 hours 45 minutes (the standard deviation is 10 minutes).

The same evaluation has been performed on the ``June 2016'' data set. Figure \ref{fig:vod201606_daily_wakeups} and \ref{fig:vod201606_daily_dormitions}, show the daily wake-up times and bedtimes, respectively. Basically, the same tendencies can be observed: wake-up times and bedtimes are shifted on holidays. However, there are some irregularities, especially within the bedtimes. On June 22, the bedtime is almost on the same level as a weekend. The mobile phone activity decrease occurred 30 minute later than on the other days of that week.

\begin{figure}
  \includegraphics[width=\linewidth]{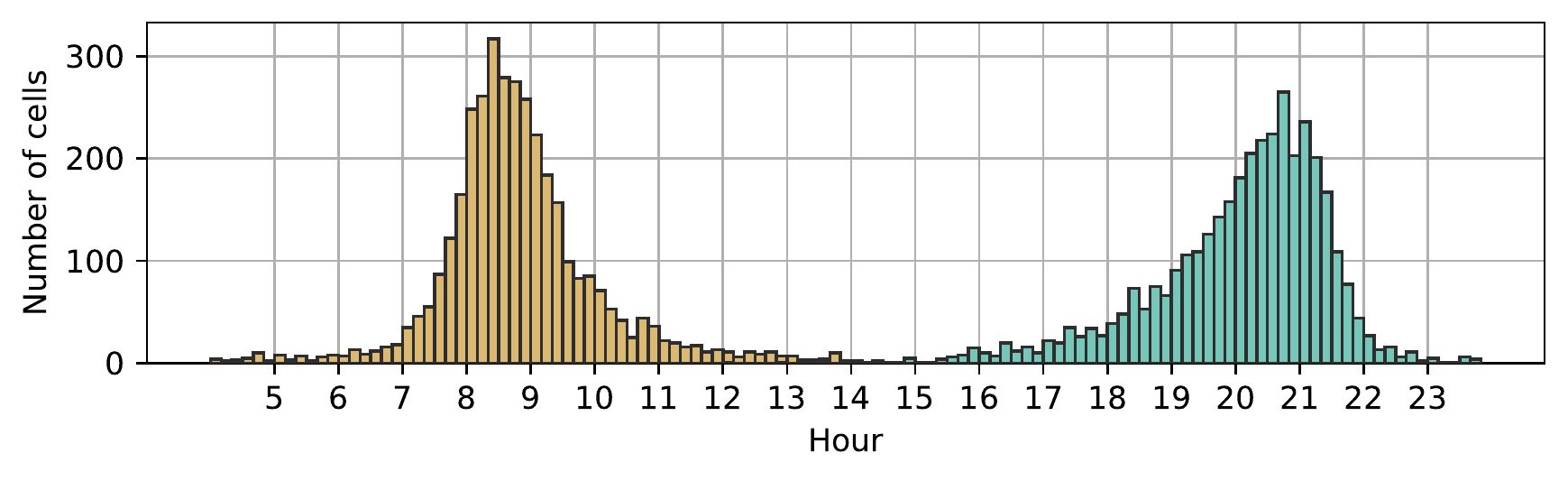}
  \caption{Inhabitant-based wake-up (brown) and bedtime (green) values of the cells, April 2017.}
  \label{fig:dweller_based_wudo_histogram}
\end{figure}

Figure~\ref{fig:dweller_based_wudo_histogram}, displays the inhabitant-based wake-up and bedtimes of the cells. This result is comparable to \cite[Figure~9]{cuttone2017sensiblesleep}, however in that case, the wake-up times ($t_{wake}$) were earlier and bedtimes were later ($t_{sleep}$).
Note that, the difference ensue from the nature of the approach. In \cite{cuttone2017sensiblesleep}, the screen-on events of the phones are considered, while in Figure~\ref{fig:dweller_based_wudo_histogram}, the active usage of the mobile phone network is aggregated for a larger area.
The ``SensibleSleep'' application \cite{cuttone2017sensiblesleep} observes offline (from a mobile phone network perspective) smartphone activity, that adjusts better to the actual \acrshort{SWC} of an individual.

\subsection{The Length of the Day}
\label{sec:day_length}

Figure \ref{fig:vod201704_daily_wakeups} and \ref{fig:vod201704_daily_dormitions}, show that people start and end their days later on the holidays. This makes perfect sense, hence they do not need to go to work, they do not need to spend time with traveling to get somewhere in time, and they probably like to rest more. But, the bedtime is shifted as well, this results that the length of the day is remains practically the same, see Figure \ref{fig:vod201704_day_lengths}.

Although, the two data sets (Section~\ref{sec:vod201606} and \ref{sec:vod201704}) were not recorded in the same year, but still covers two different months of a year (April and June).
Considering that a mid-spring month is well represented by the ``April 2017'' dataset, and an early summer month by the ``June 2016'' dataset (for Budapest), the differences between the wake-up time, bedtime, and the day length can be compared between two seasons.

Figure~\ref{fig:vod201606_daily_wakeups}, shows a small increase in the wake-up times, during the second half of the month. Although, that 10-minute increase in the averages cannot be considered significant, it may reflect the end of the school term (June 15). During the intersession, the time schedule of the public transport services is adjusted. For many lines, the headway is increased on workdays, while for some lines, it is decreased, especially on weekends. In this way, the intersession affects the whole transportation system of Budapest, and even its agglomeration to some extent.

\begin{figure}
    \begin{subfigure}[t]{0.49\linewidth}
        \includegraphics[width=\linewidth]{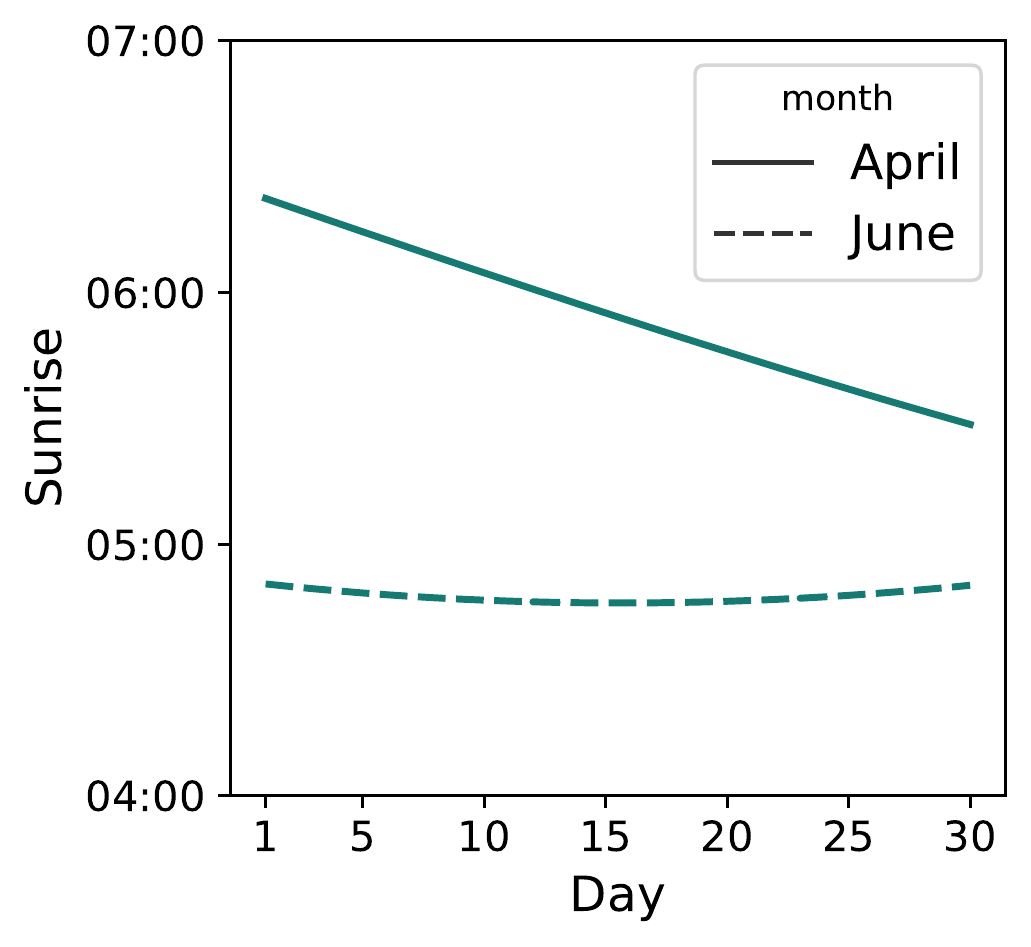}
        \captionsetup{position=bottom,justification=centering}
        \caption{}
        \label{fig:sunrise}
    \end{subfigure}
    \hfill
    \begin{subfigure}[t]{0.49\linewidth}
        \includegraphics[width=\linewidth]{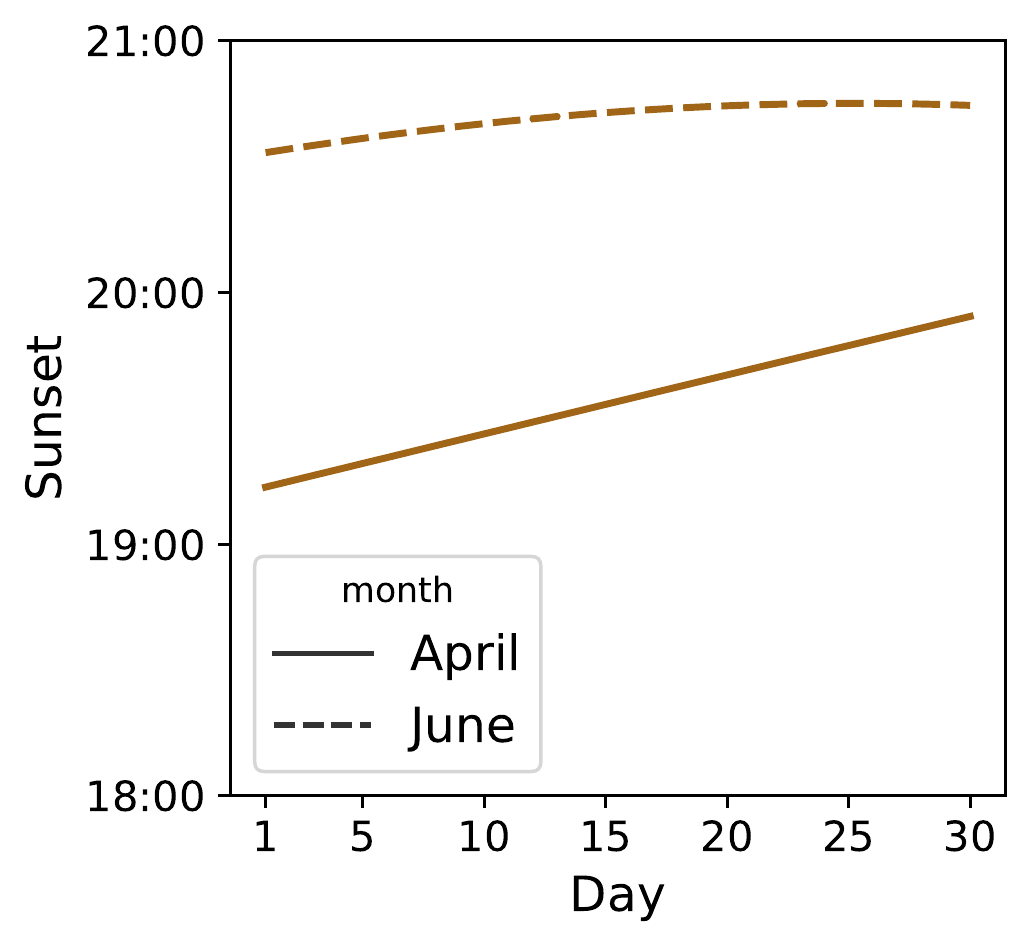}
        \captionsetup{position=bottom,justification=centering}
        \caption{}
        \label{fig:sunset}
    \end{subfigure}
    \caption{Sunrise (\textbf{a}) and sunset (\textbf{b}) times of June 2016 and April 2017 projected to the same figure to highlight the differences between the seasons, using data from \cite{visualcrossing}.
    }
    \label{fig:sunrise_sunset}
\end{figure}

Until the summer solstice (June 20 in 2016 \cite{espenak2018solstices}), the days are getting longer, but can the longer daylight have en effect to mobile phone network activity?
As a reference, astronomical information (sunrise and sunset) has obtained from \url{www.visualcrossing.com} \cite{visualcrossing} for Budapest.
Figure~\ref{fig:sunrise_sunset}, shows the difference between the sunrises and the sunsets during the two observation periods of the data sets, projected to the same figure. The dashed lines display the values of June, and the solid lines display the values of April. As the summer solstice is in June, the differences during the month are negligible, but in April, the differences between the beginning and the end of the month are much more significant.

So, in June, the Sun rises earlier and sets later than in April, consequently lighter period of days are longer. June 15 is more than 2-hour longer than April 15, using the astronomical definitions of the sunrise and sunset.
According to the calculated day lengths, the average workday length is 12 hours in the ``June 2016'' data set, and 10 hours and 40 minutes in the ``April 2017'' data set, resulting a difference of the 80 minutes.

Although, this is less than the astronomical difference, but the Sun rises very early in the morning, when people are still sleeping. It may be more practical to compare the results with the difference of the sunsets. As, the people do not wake up earlier just to organize an activity before work, but may do after work if it is still bright and the weather is good. The Sun sets at 19:33 on April 15 and at 20:42 on June 15, that is a 69-minute difference, that is more comparable to the calculated values. The average workday bedtime values are 19:43 and 20:47, respectively.
This finding --- the day lengths are longer in June --- is in good agreement with \cite{monsivais2017seasonal}, where the seasonal influence of the daylight was examined via the length resting period.

The wake-up times are 8:57 and 8:44 in average (workdays), that show a slight decrease in the summer. Figure~\ref{fig:vod201606_daily_wakeups}, shows slightly later wake-up times in the second half of the month. The average workday wake-up time in the first half of June 2016 is 8:39, and 8:50 in the second half. As mentioned in Section~\ref{sec:dweller_based_wakeup}, it might be caused by the end of the school term.

\subsection{Area-based Approach}
\label{sec:area_based_wakeup}

\begin{figure}
  \begin{subfigure}[th]{0.495\linewidth}
    \includegraphics[width=1\linewidth]{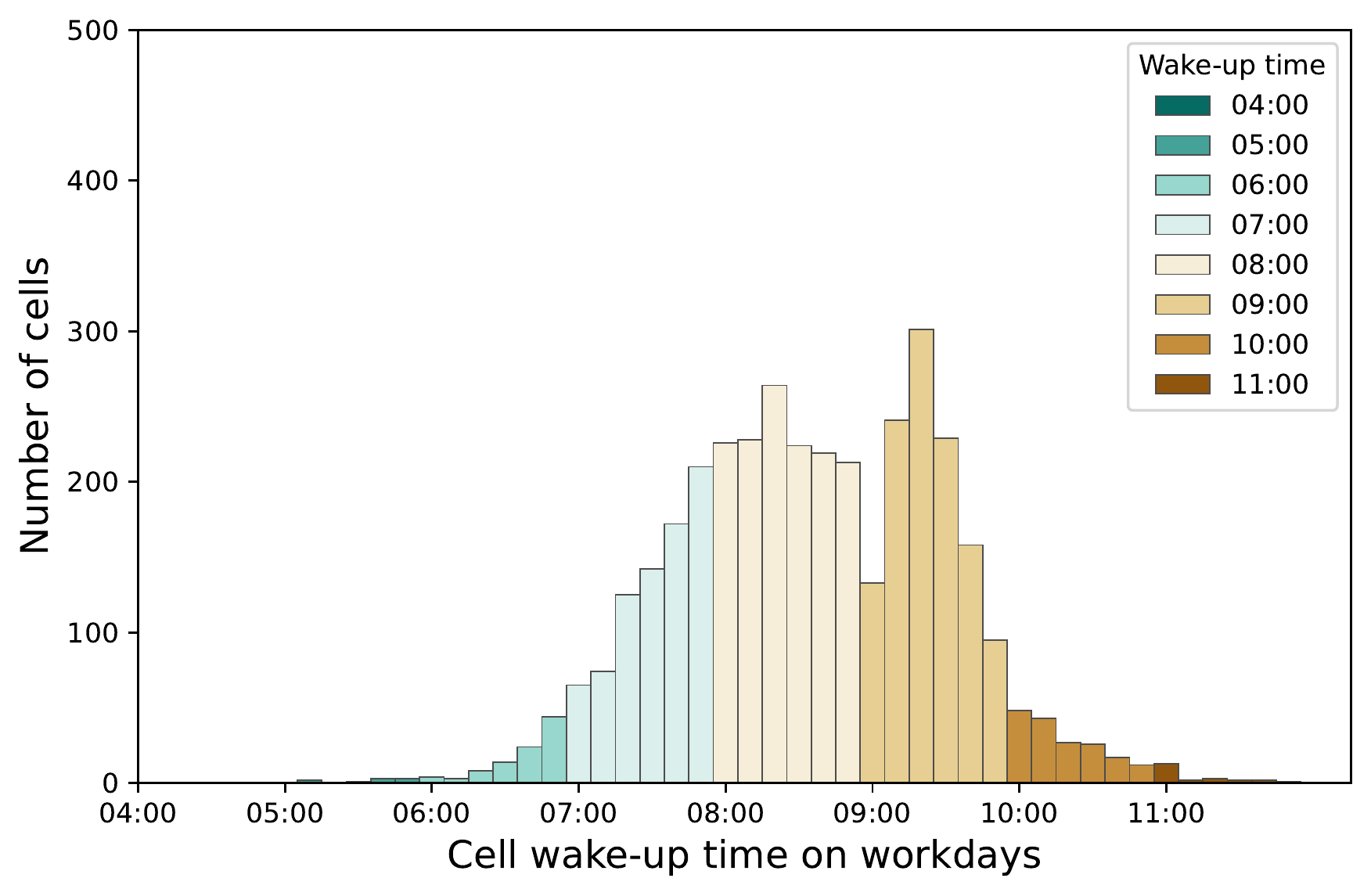}
    \captionsetup{position=bottom,justification=centering}
    \caption{}
    \label{fig:cell_wakeup_weekday_histogram}
  \end{subfigure}
  \begin{subfigure}[th]{0.495\linewidth}
    \includegraphics[width=1\linewidth]{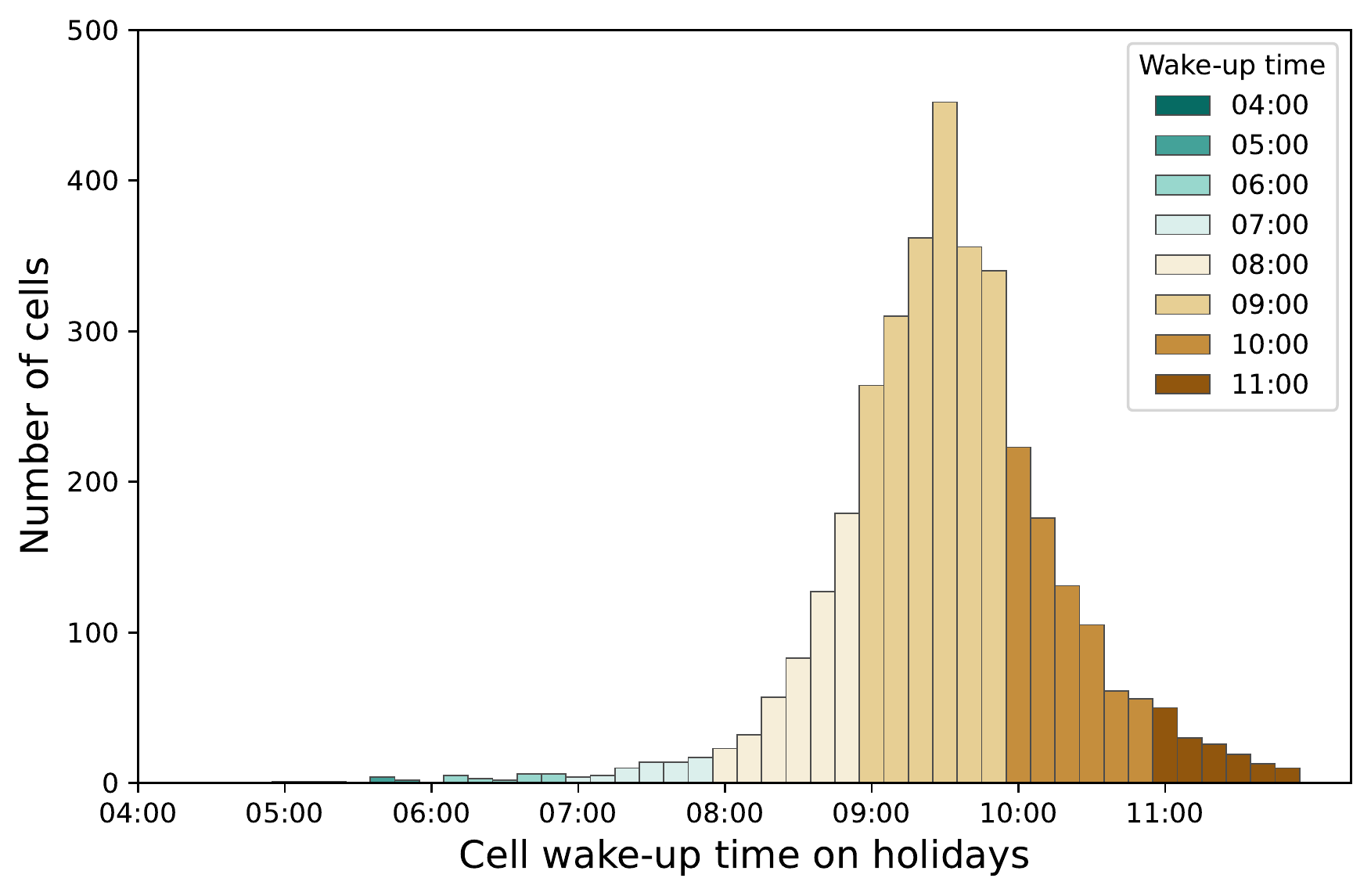}
    \captionsetup{position=bottom,justification=centering}
    \caption{}
    \label{fig:cell_wakeup_weekend_histogram}
  \end{subfigure}
  \vfill
  \vspace{.75em}
  \begin{subfigure}[th]{0.495\linewidth}
    \includegraphics[width=1\linewidth]{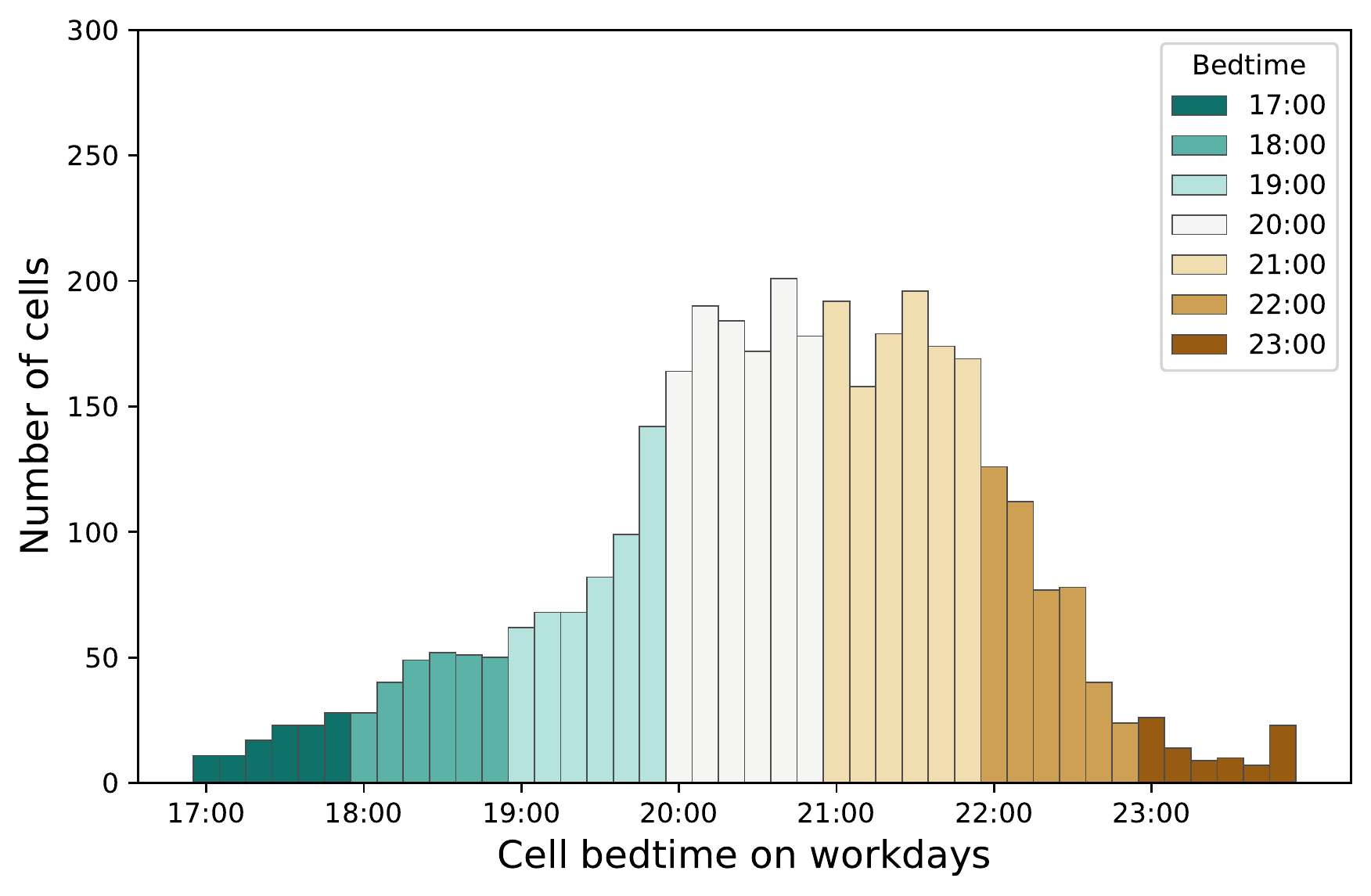}
    \captionsetup{position=bottom,justification=centering}
    \caption{}
    \label{fig:cell_bedt_weekday_histogram}
  \end{subfigure}
  \begin{subfigure}[th]{0.495\linewidth}
    \includegraphics[width=1\linewidth]{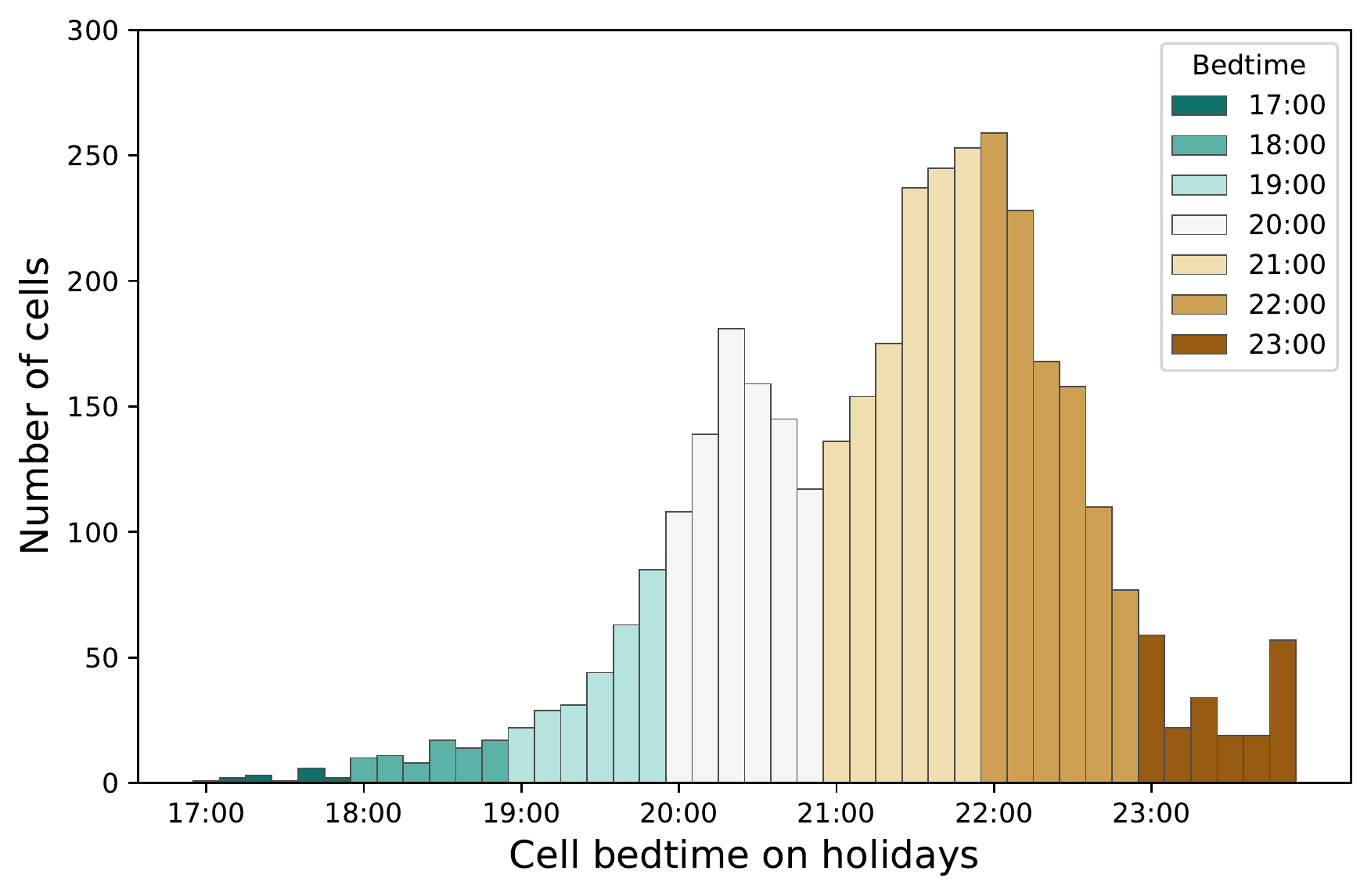}
    \captionsetup{position=bottom,justification=centering}
    \caption{}
    \label{fig:cell_bedt_weekend_histogram}
  \end{subfigure}
  \caption{Cell based wake-up (\textbf{a}, \textbf{b}) and bedtime (\textbf{c}, \textbf{d}) distribution for workdays and holidays, respectively, in the ``April 2017'' dataset.}
  \label{fig:cell_wakeup_histograms}
\end{figure}

Based on the cell/area activity, the wake-up, and the bedtime has been determined for every cell. Figure \ref{fig:cell_wakeup_weekday_histogram} and \ref{fig:cell_bedt_weekday_histogram}, show the distribution for workdays,
and Figure \ref{fig:cell_wakeup_weekend_histogram} and \ref{fig:cell_bedt_weekend_histogram}, show distribution for holidays, respectively.
The results clearly show that the usage of the mobile phone network intensifies and reduces later on the holidays, indicating the people wake up and go to sleep later when they do not work.

\begin{figure*}
    \includegraphics[width=1\linewidth]{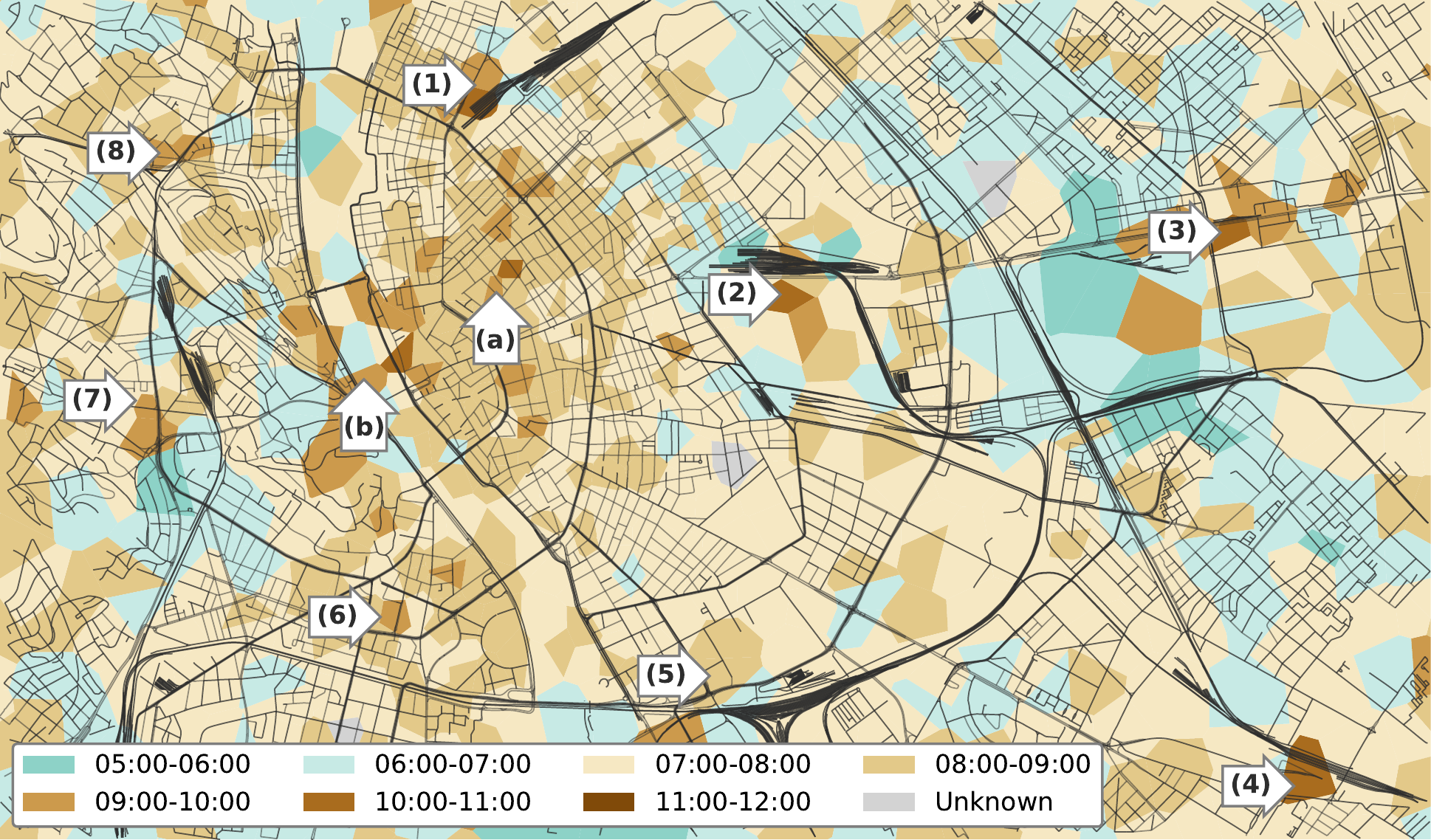}
    \caption{Spatial distribution of the cell-based wake-up times, malls (1-8) opens at 10:00.}
    \label{fig:cell_wakeup_malls}
\end{figure*}

\begin{figure*}
    \includegraphics[width=1\linewidth]{figures/vod201704/site_bedtime_brbg}
    \caption{Spatial distribution of the cell-based bedtimes, aggregated to sites.}
    \label{fig:site_bedtime}
\end{figure*}

Figure~\ref{fig:cell_wakeup_malls}, shows the spatial distribution of the cell-based wake-up times. The Voronoi polygons, representing the mobile phone cells, are colored by the calculated ``wake-up'' times. The map clearly shows some extrema, where the times are around 10:00. Extrema 1-8 are all malls
(WestEnd City Center (1), Arena Mall (2), Árkád (3), KÖKI Terminál (4), Lurdy Ház (5), Allee (6), MOM Park (7) and Mammut (8).), that uniformly open at 10:00, though 5 and 8 partly serves as office buildings.

Figure~\ref{fig:site_bedtime}, shows the bedtime values of the sites, on workdays. As the Figure~\ref{fig:cell_bedt_weekday_histogram}, illustrates in respect of cells, the bedtime is usually between 20:00 and 22:00. There are, however, a some sites with later bedtime, a few of them are denoted in the figure. Marker 1 at the party district (Appendix~\ref{app:party_district}).
In the site at marker 2, does not have any distinctive object, that could explain this result. However, East of that site, there is a beer factory, that might have notable activity in the evening --- compared to the neighborhood --- and the distortion of the Voronoi tessellation could have resulted this late bedtime.
At marker 3, there is a student's hostel and, in the neighborhood, there are some sport and concert venues.

\subsection{Working Hours}
\label{sec:working_hours}

\begin{figure}
    \begin{subfigure}[t]{0.4925\linewidth}
        \includegraphics[width=\linewidth]{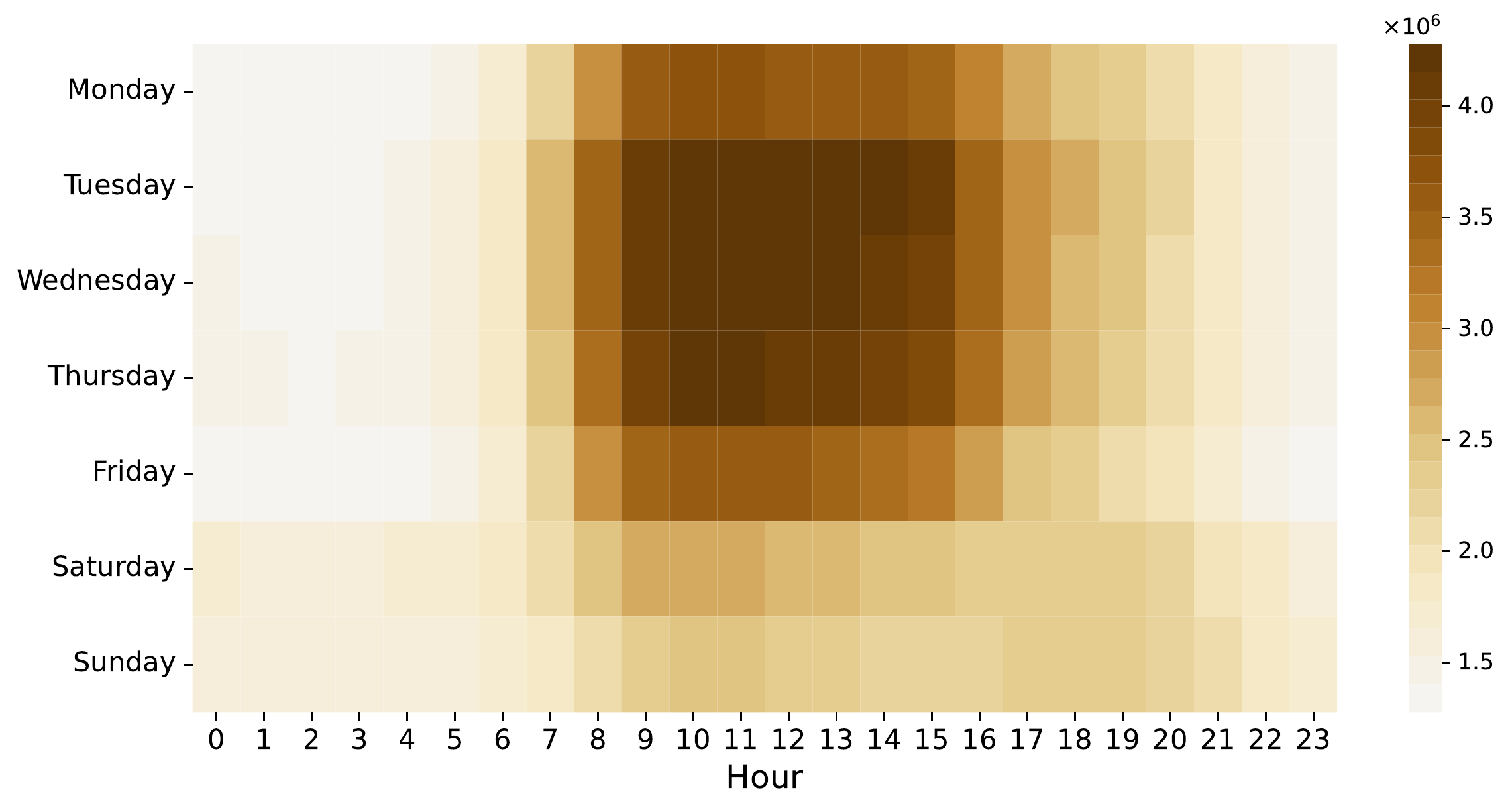}
        \captionsetup{position=bottom,justification=centering}
        \caption{}
        \label{fig:day_hour_work_activity}
    \end{subfigure}
    \hfill
    \begin{subfigure}[t]{0.4925\linewidth}
        \includegraphics[width=\linewidth]{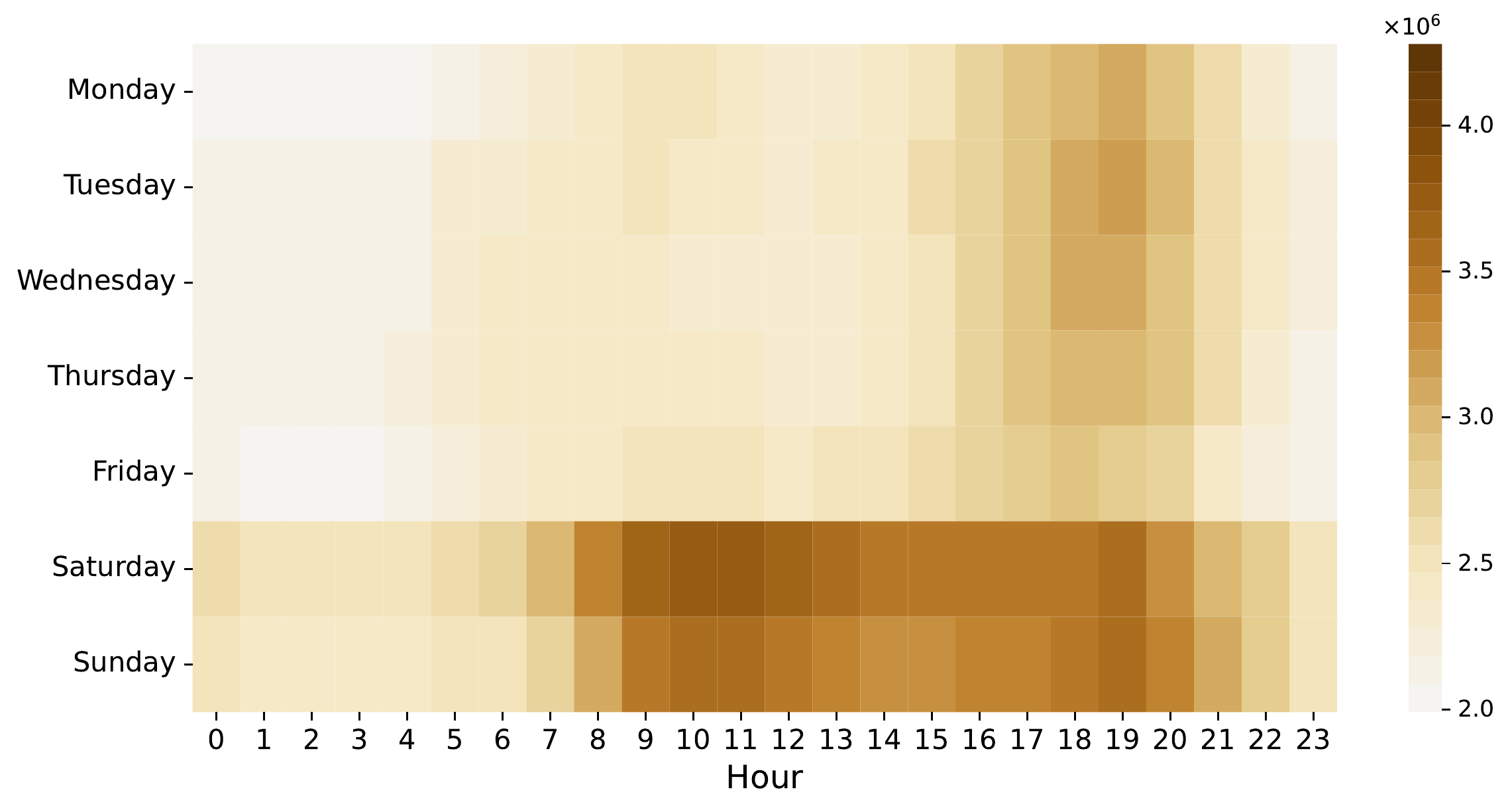}
        \captionsetup{position=bottom,justification=centering}
        \caption{}
        \label{fig:day_hour_home_activity}
    \end{subfigure}
    \caption{The mobile phone activity distribution by days of week and hours based on the ``April 2017'' data set, for the work places (\textbf{a}) and the homes (\textbf{b}).
    }
    \label{fig:day_hour_work-home_activity}
\end{figure}

Figure~\ref{fig:day_hour_work-home_activity}, shows the activity distribution by days of week and hours, based on the ``April 2017'' data set, separating the workplace (\ref{fig:day_hour_work_activity}) and the home (\ref{fig:day_hour_home_activity}) activity, but using the same scale. At the workplace, most of the activity recorded during the working hours on workdays, as expected. The activity increases fast in the mornings, but decreases more slowly in the afternoon.
The home activity was the mostly clustered on the weekends, but after the work hours, in the evening, had a notable activity peak. The home activity before the working hours does not seem so significant.

\begin{figure}
    \begin{subfigure}[t]{\linewidth}
        \includegraphics[width=\linewidth]{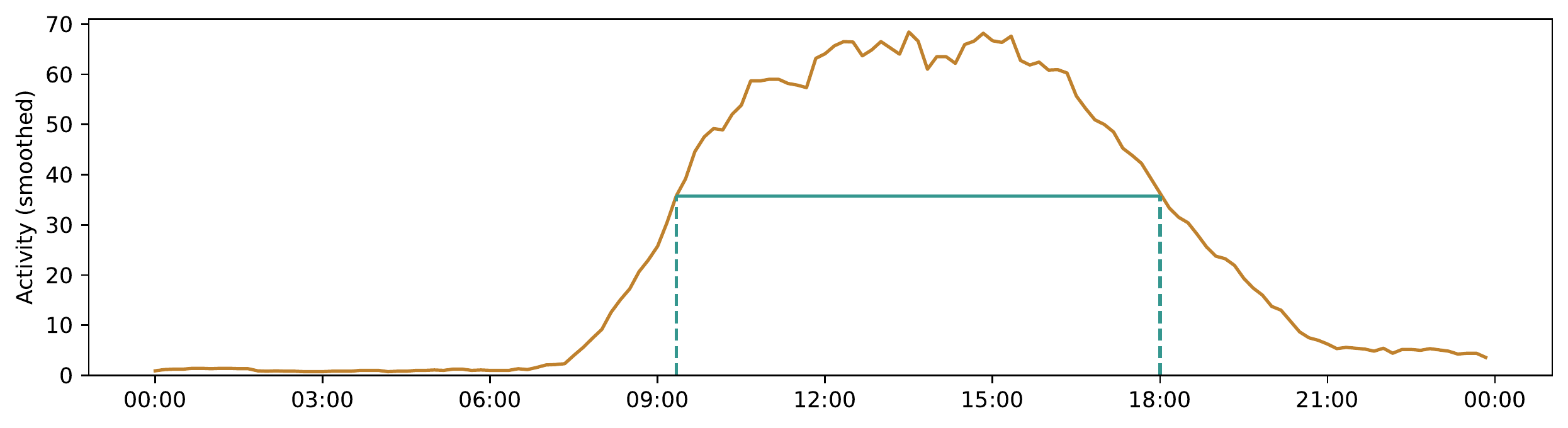}
        \captionsetup{position=bottom,justification=centering}
        \caption{}
        \label{fig:site_work_exclusive}
    \end{subfigure}
    \vfill{}
    \vspace{0.5em}
    \begin{subfigure}[t]{\linewidth}
        \includegraphics[width=\linewidth]{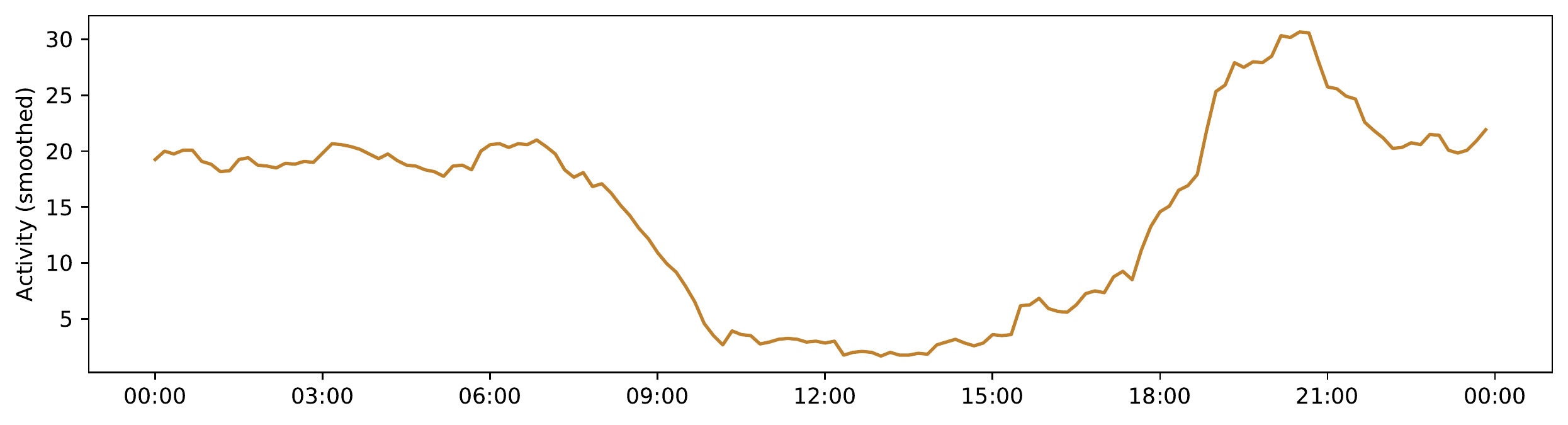}
        \captionsetup{position=bottom,justification=centering}
        \caption{}
        \label{fig:site_home_exclusive}
    \end{subfigure}
    \caption{Activity of the workers (\textbf{a}) and the inhabitants (\textbf{b}) of a selected site, on a selected day.}
    \label{fig:work-home_exclusive_sample}
\end{figure}

This procedure can be applied for every cell (or group of cells, like sites), then the activity of the workers and the inhabitants will represent the workplace and home activity tendencies.
Note that, a cell can have both workers and inhabitants, so every cell have two aspects.
Figure~\ref{fig:site_work_exclusive} and \ref{fig:site_home_exclusive}, illustrates the activity of the subscribers who work and live a selected site, respectively. Figure~\ref{fig:site_work_exclusive}, also demonstrates the concept of Figure~\ref{fig:wakeup_calc}, using actual data.

As expected, Figure~\ref{fig:site_work_exclusive} and \ref{fig:site_home_exclusive} is in accord with Figure~\ref{fig:day_hour_work_activity} and Figure~\ref{fig:day_hour_home_activity}. Workers' activity increases in the morning and decreases late in the afternoon, whereas the inhabitants' activity decreases in the morning and increases late in the afternoon and reaches its peak in the evening. The two aspect of the same site has the opposite tendency.

Applying the same, inhabitant-based approach as in the case of the wake-up time (Figure~\ref{fig:wakeup_calc}) to the workers' activity, it is possible to detect the positive and the negative edge of the activity curve. The positive edge could indicate the start of the working hours, and the negative edge could indicate the end of the working hours in a given cell (or site). Moreover, the difference of the two times can determine the length of the working hours.

\begin{figure}
  \centering
    \includegraphics[width=\linewidth]{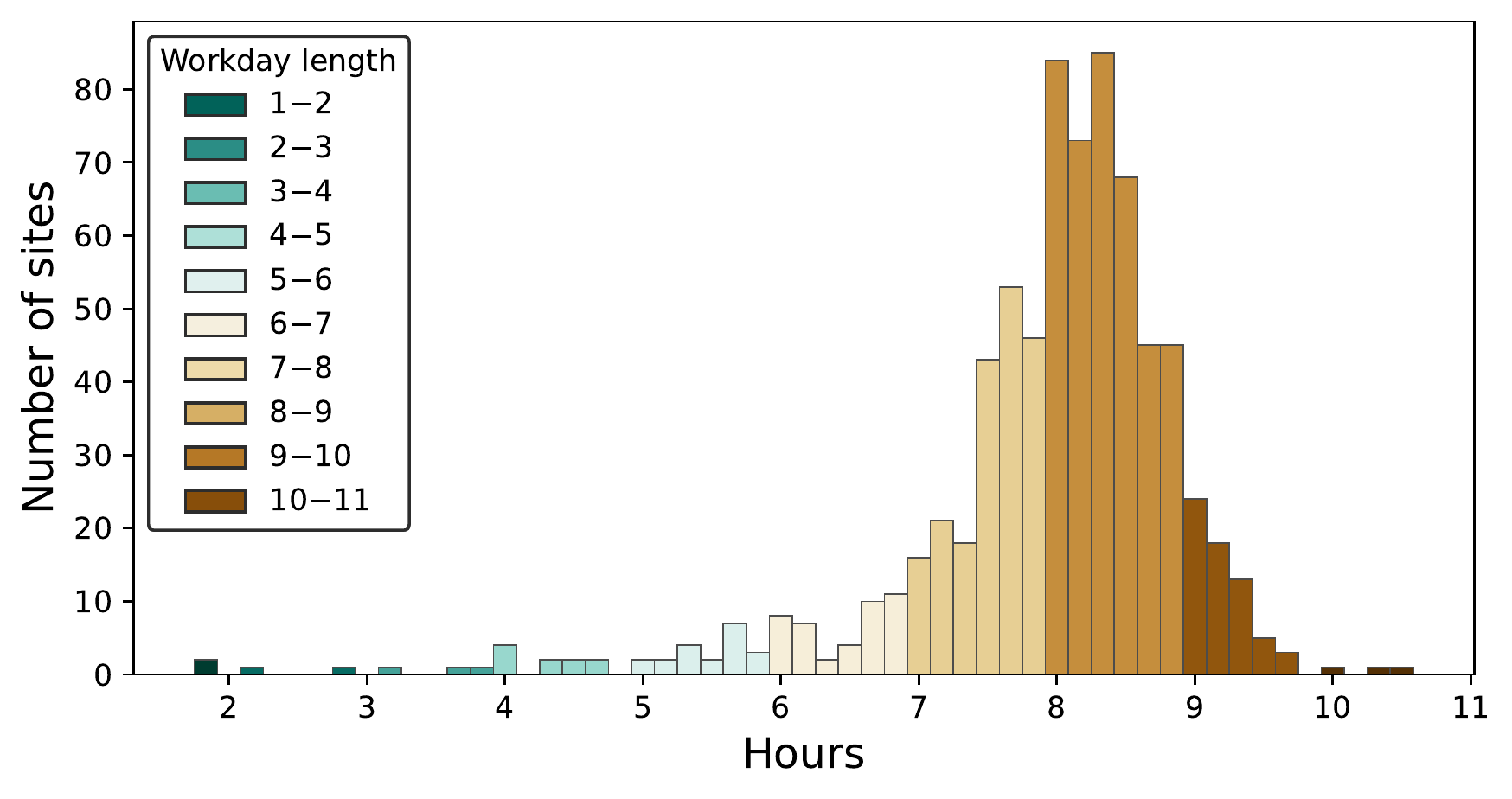}
    \caption{Distribution of workday length in sites.}
    \label{fig:workday_length_site}
\end{figure}

In most of the sites, the working hours are about 8 hours (Figure~\ref{fig:workday_length_site}), just as expected. In the rest of the sites, --- especially where the working hour is less than 7 hours or over 9 and a half hours --- the mobile phone activity proved to be so low during the working hours, that the results cannot be considered appropriate. Nevertheless, the mobile phone network reflects the length of the working hours.

The work hours do not necessarily start and end at the same time in every workplace. Are there any differences in this regard, from a mobile phone perspective?

Figure~\ref{fig:workday_length_site}, shows the average workday length of the sites. The workday length in most of them are around eight hours, that agrees with the practice, in Hungary.
Moreover, the distribution of the working hour lengths also verifies the workplace detection approach, summarized in Section~\ref{sec:home_work}.
Although, there are some sites with very low (less than seven hours) or high (more than nine and a half hours), but these sites have very few activities during the observation period: less than 2\% of the total activity occurs in these sites.

\begin{figure}
    \begin{subfigure}[t]{0.49\linewidth}
        \includegraphics[width=\linewidth]{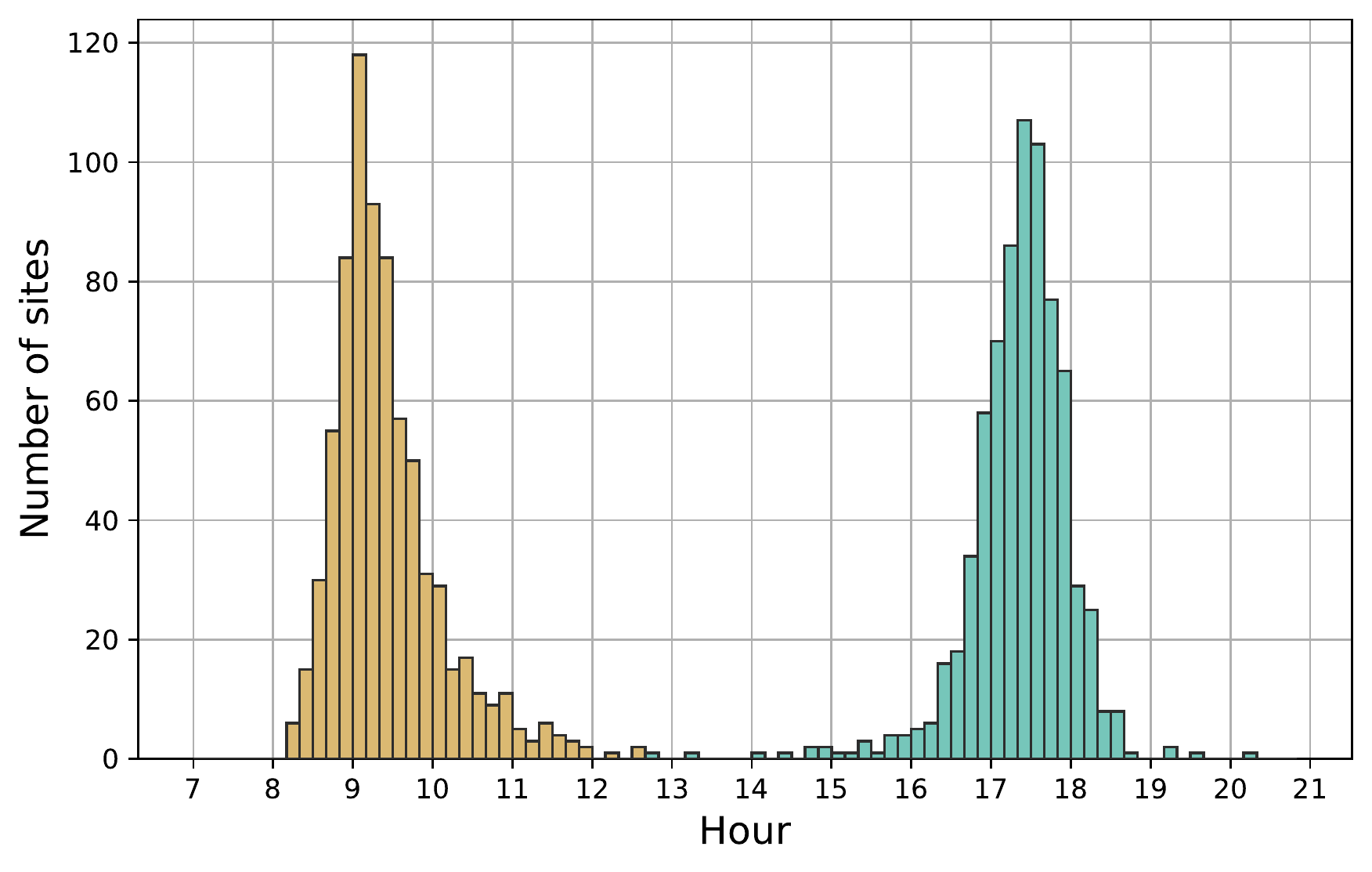}
        \captionsetup{position=bottom,justification=centering}
        \caption{}
        \label{fig:working_hours_start_end_histogram}
    \end{subfigure}
    \hfill{}
    \begin{subfigure}[t]{0.49\linewidth}
        \includegraphics[width=\linewidth]{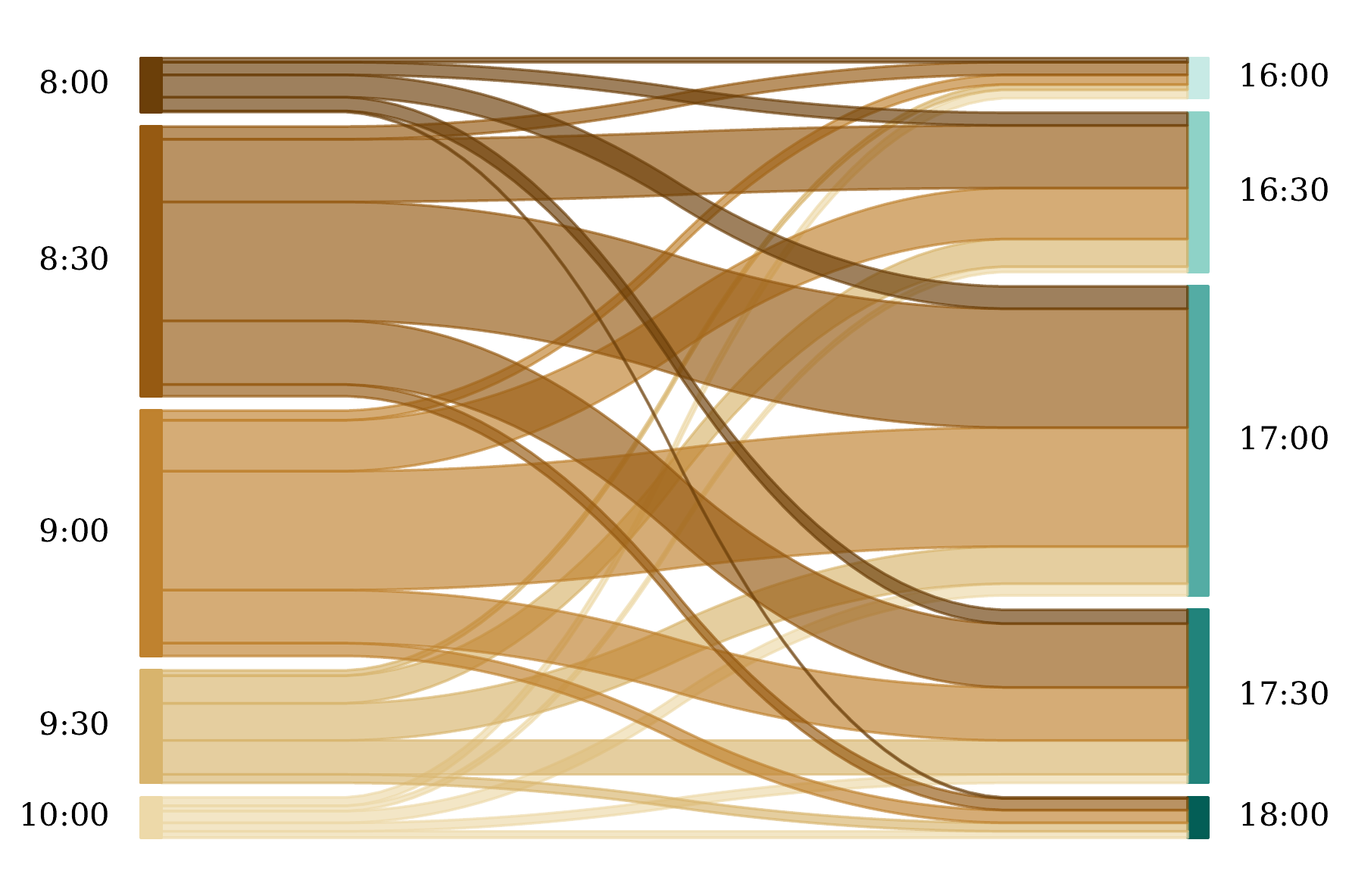}
        \captionsetup{position=bottom,justification=centering}
        \caption{}
        \label{fig:workinghours_sankey}
    \end{subfigure}
    \caption{The distribution of the starting (brown) and the ending (green) of the working hours, in sites (\textbf{a}); and the connection between the starting and the ending times (\textbf{b}).}
    \label{fig:working_hours}
\end{figure}

The length of the working hours is usually around eight hours in most of the sites, but there might be differences at the beginning and the end of the labor-time.
Figure~\ref{fig:workinghours_sankey}, shows the five most frequent beginning and ending labor-time (defined as the positive and negative edge of the workplace activity curve of the site) and the connection between them, using a type of Sankey diagram.
It is clear, that in most of the sites, the mobile phone network activity increases between 8:30 and 9:30.
The most notable group
That is not surprising, considering he lunch break.
Note that, the observed time values may have a delay compared to the actual start of the work. As an employee may not actively use the mobile phone network as soon as they arrived to their workplace (or when they left it).

\subsection{In Respect of Mobility}

\begin{figure}
    \begin{subfigure}[t]{0.49\linewidth}
        \includegraphics[width=1\linewidth]{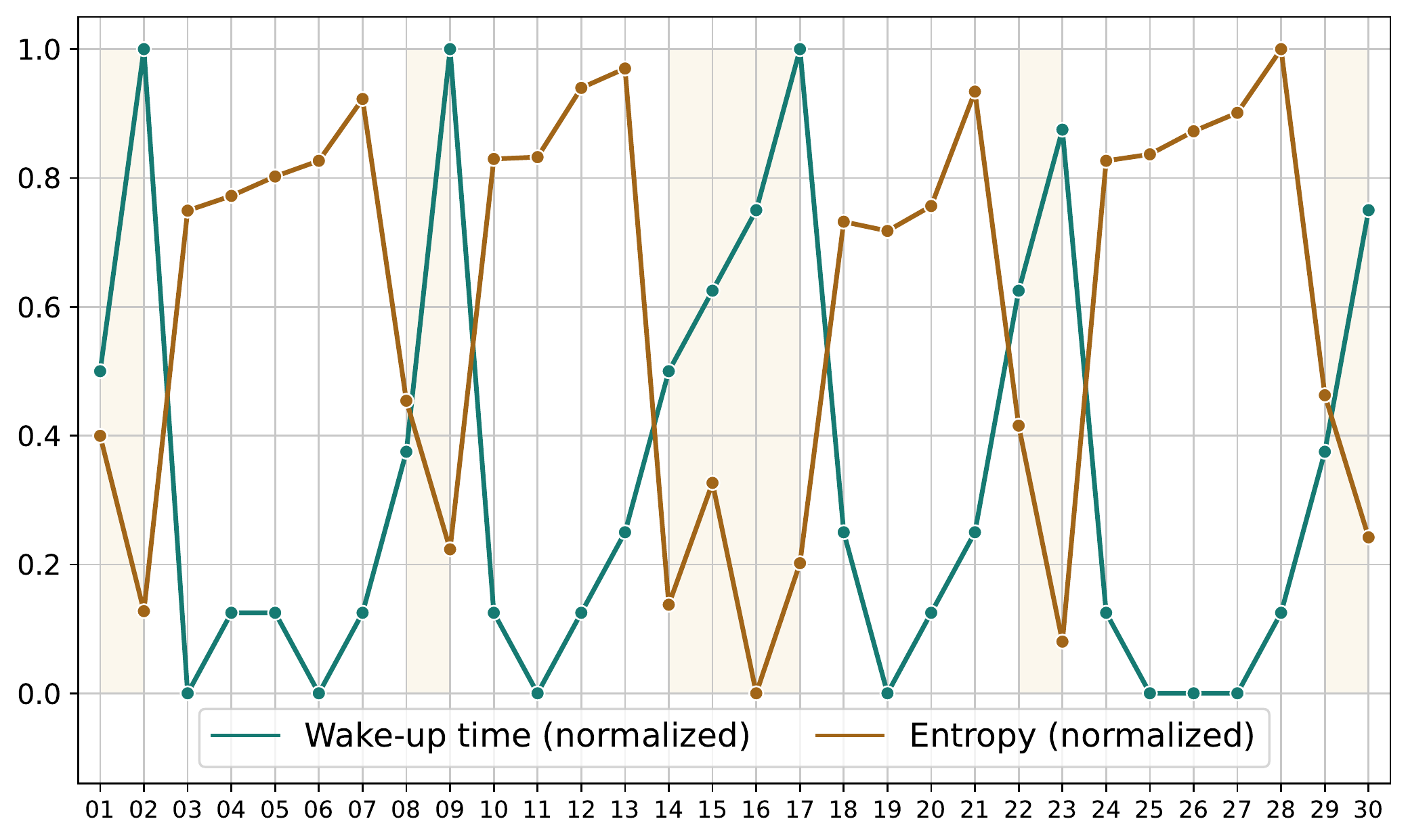}
        \captionsetup{position=bottom,justification=centering}
        \caption{}
        \label{fig:entropy_vs_wakeup}
    \end{subfigure}
    \hfill
    \begin{subfigure}[t]{0.49\linewidth}
      \includegraphics[width=1\linewidth]{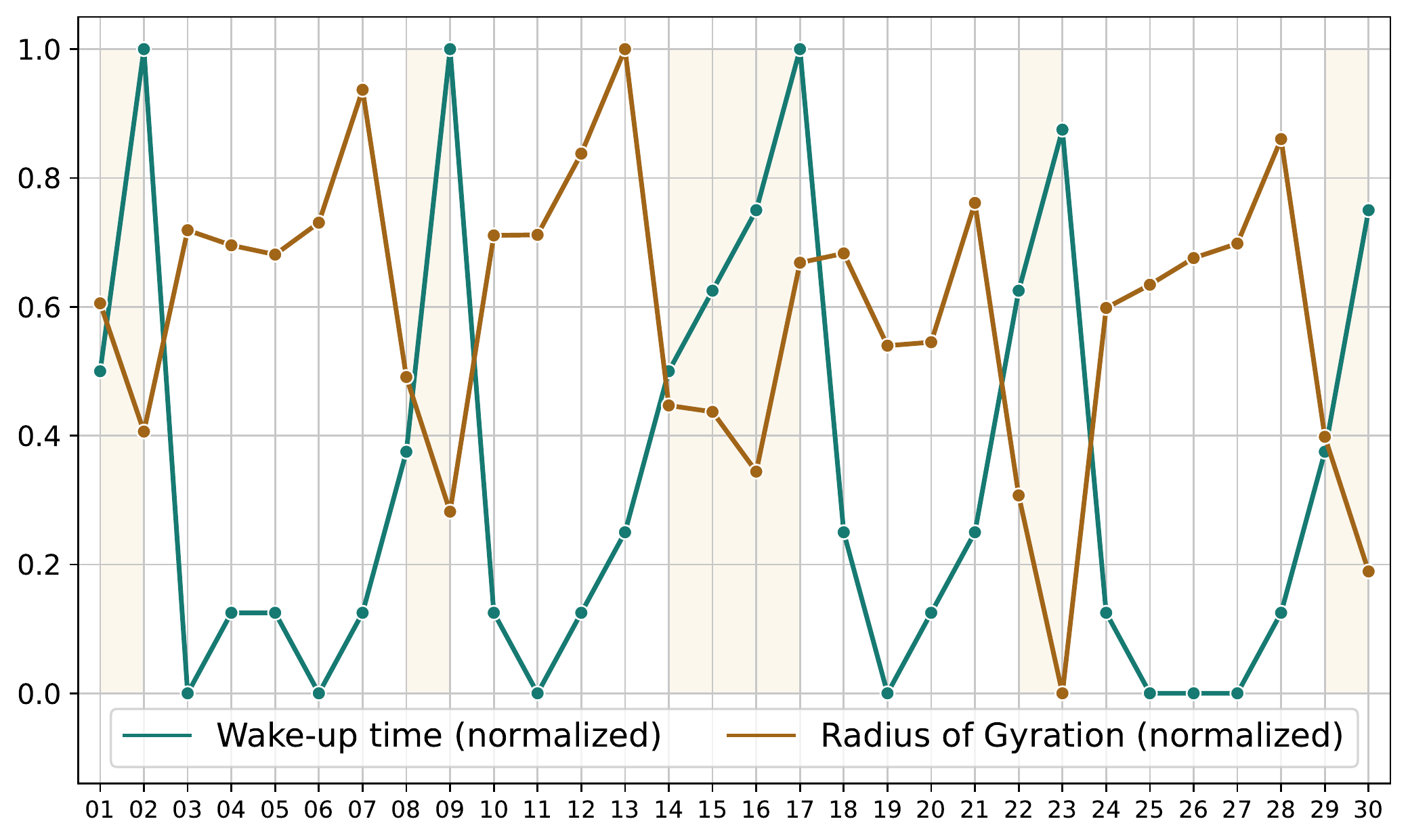}
      \captionsetup{position=bottom,justification=centering}
      \caption{}
      \label{fig:gyration_vs_wakeup}
    \end{subfigure}
    \vfill
    \vspace{0.5em}
    \begin{subfigure}[t]{0.49\linewidth}
        \includegraphics[width=1\linewidth]{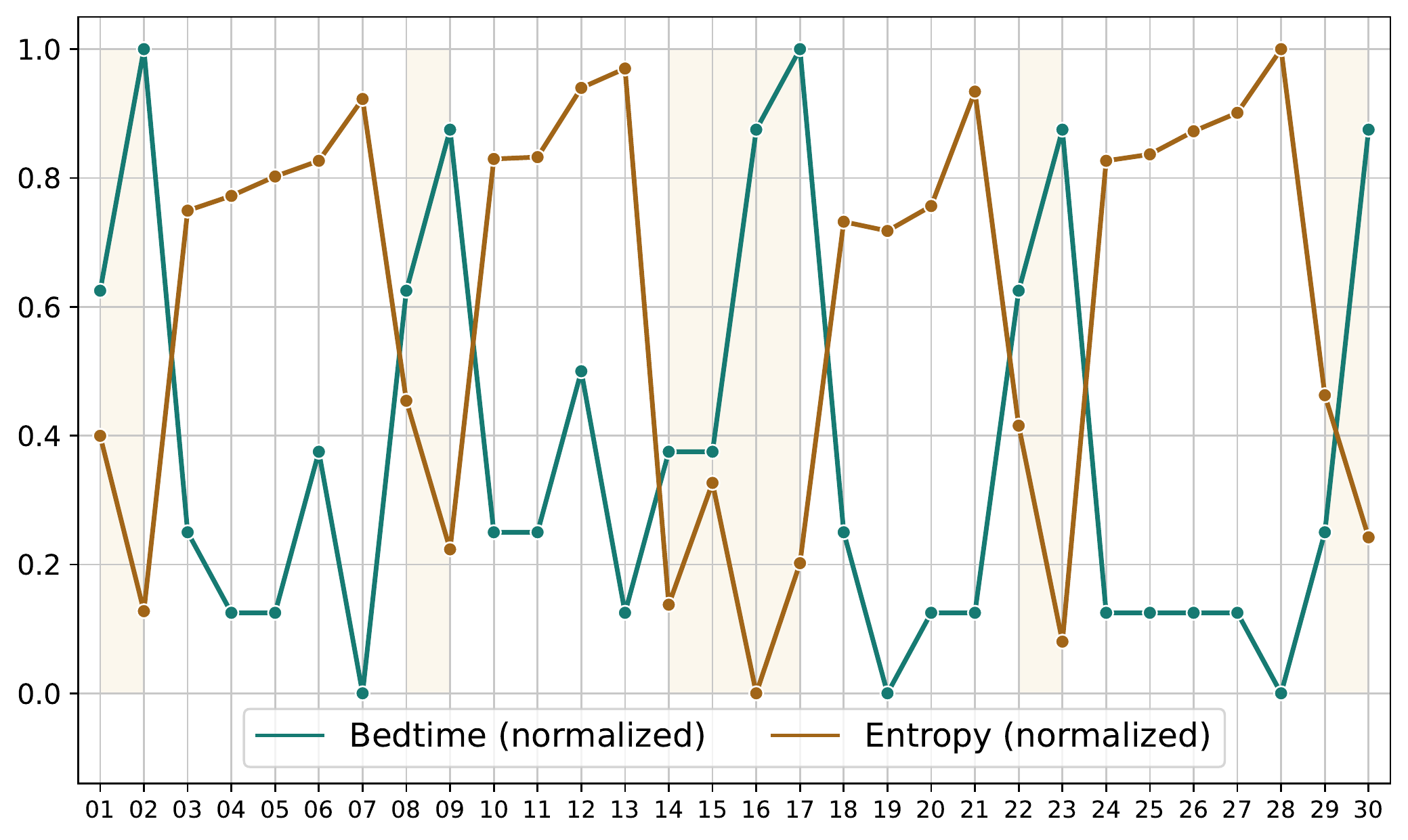}
        \captionsetup{position=bottom,justification=centering}
        \caption{}
        \label{fig:entropy_vs_bedtime}
    \end{subfigure}
    \hfill
    \begin{subfigure}[t]{0.49\linewidth}
      \includegraphics[width=1\linewidth]{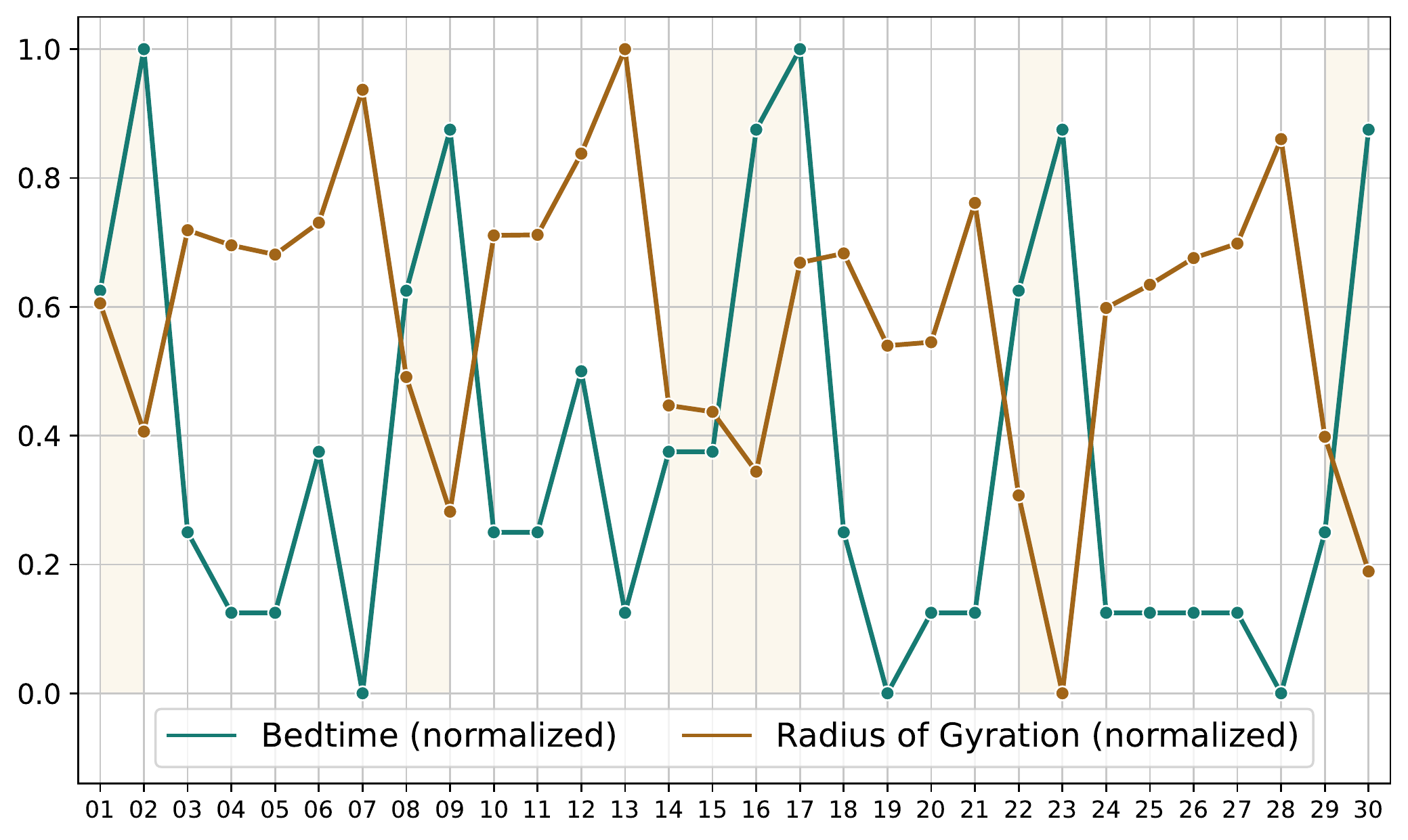}
      \captionsetup{position=bottom,justification=centering}
      \caption{}
      \label{fig:gyration_vs_bedtime}
    \end{subfigure}
    \caption{Normalized, inhabitant-based wake-up times in contrast of the normalized daily Entropy (\textbf{a}) and Radius of Gyration (\textbf{b}). Pearson's Rs are \num{\DailyEntWuPearsonR} and \num{\DailyGyrWuPearsonR}, respectively.
    Second row: Normalized, inhabitant-based bedtimes in contrast of the normalized daily Entropy (\textbf{c}) and Radius of Gyration (\textbf{d}). Pearson's Rs are \num{\DailyEntDoPearsonR} and \num{\DailyGyrDoPearsonR}, respectively.}
    \label{fig:mobility_vod201704}
\end{figure}

In \cite{pinter2021evaluating}, we also showed that the mobility has a weekly seasonality. As Figure~\ref{fig:daily_wakeup_dormition_and_daylength} shows, the wake-up time also have a seasonality.
To compare it with the mobility indicators, described in Section~\ref{sec:mobility_metrics}, the daily wake-up times and the mobility metrics were normalized, using min-max feature scaling. Figure~\ref{fig:mobility_vod201704}a and Figure~\ref{fig:mobility_vod201704}b, displays the normalized Entropy and the normalized Radius of Gyration --- without the non-phone devices --- in contrast of the wake-up times, respectively.
As the mobility indicators are determined per subscriber, the inhabitant-based version of the wake-up time is used.

During workdays, both the Entropy and the Radius of Gyration were high, but the wake-up time was low. On holidays, it is on the contrary, the wake-up times were higher, the mobility indicators were lower.
This is not surprising, hence people tend to wake up earlier on workdays to go to work, thus people usually need to travel to their workplace.
On holidays, it is common to spend more time at home, that also reduces the mobility values.
Figure~\ref{fig:mobility_vod201704}, visualizes this clearly, especially in the case of Entropy.
Numerically (Pearson's R), the correlation between the wake-up times and the mobility indicators (Entropy, Radius of Gyration) are \num{\DailyEntWuPearsonR} and \num{\DailyGyrWuPearsonR}, respectively.

Bedtimes show a similar trend (Figure~\ref{fig:mobility_vod201704}c and Figure~\ref{fig:mobility_vod201704}d) in contrast of the two mobility metrics (Entropy and Radius of Gyration), however, the correlations numerically are not that strong: \num{\DailyEntDoPearsonR} and \num{\DailyGyrDoPearsonR}, respectively.
This might have caused by leisure activities, after work. While people may go straight to work in the morning, they do not necessarily go home right after work.

\subsection{In Respect of Socioeconomic Status}

In an earlier work \cite{pinter2021evaluating}, we demonstrated correlation between the mobility customs, the distance of the home and work locations and the socioeconomic status.
We found that people who live in less expensive parts of Budapest, tend to travel more to their workplace. The larger distance should indicate longer travel times and earlier wake-up times, as the work mostly starts at the same time in the morning.
Lotero et al. previously found that ``rich people do not rise early'' \cite{lotero2016rich}, analyzing two cities of Colombia.

\begin{figure}[th]
    \begin{subfigure}[t]{0.485\linewidth}
        \includegraphics[width=1\linewidth]{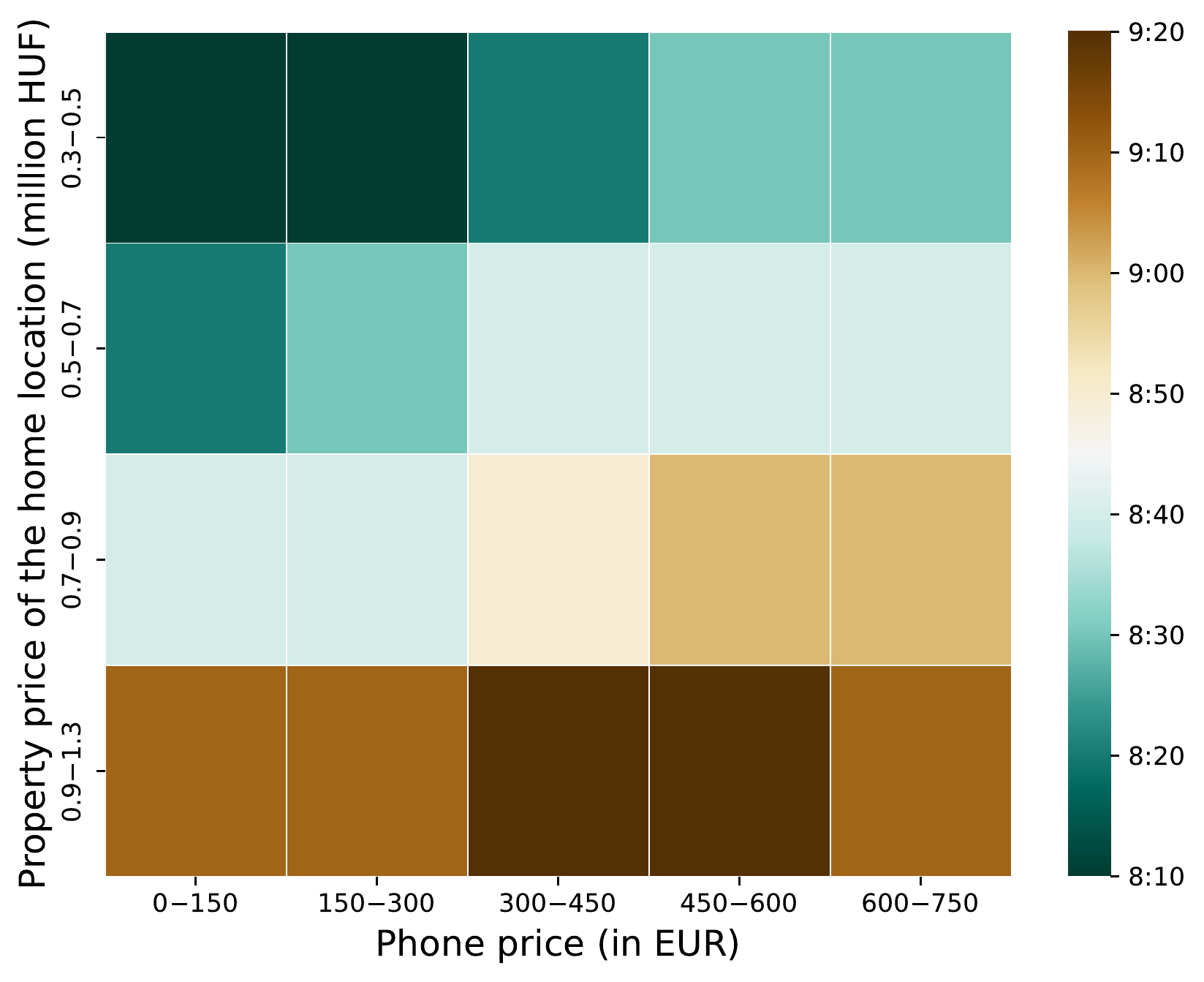}
        \captionsetup{position=bottom,justification=centering}
        \caption{}
        \label{fig:hp_pp_wuwd}
    \end{subfigure}
    \hfill
    \begin{subfigure}[t]{0.4975\linewidth}
      \includegraphics[width=1\linewidth]{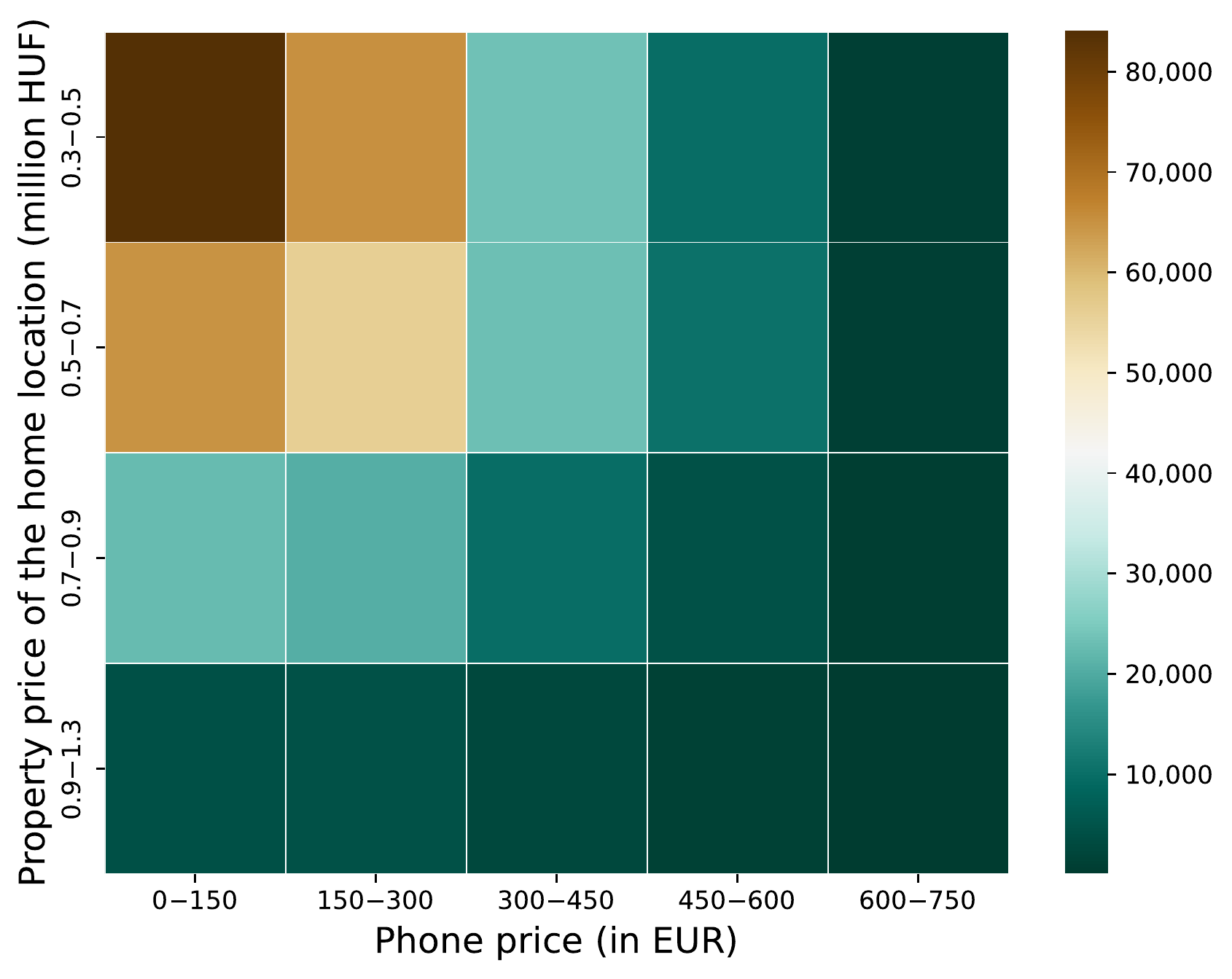}
      \captionsetup{position=bottom,justification=centering}
      \caption{}
      \label{fig:hp_pp_wuwd_cardinality}
    \end{subfigure}
    \caption{The inhabitant-based wake-up times in the socioeconomic categories, based on the property price of the home location and mobile phone price (\textbf{a}), and the number of the subscribers in each category (\textbf{b}).}
    \label{fig:homeprice_phoneprice_wakeup}
\end{figure}

\begin{figure}[th]
    \begin{subfigure}[t]{0.485\linewidth}
        \includegraphics[width=1\linewidth]{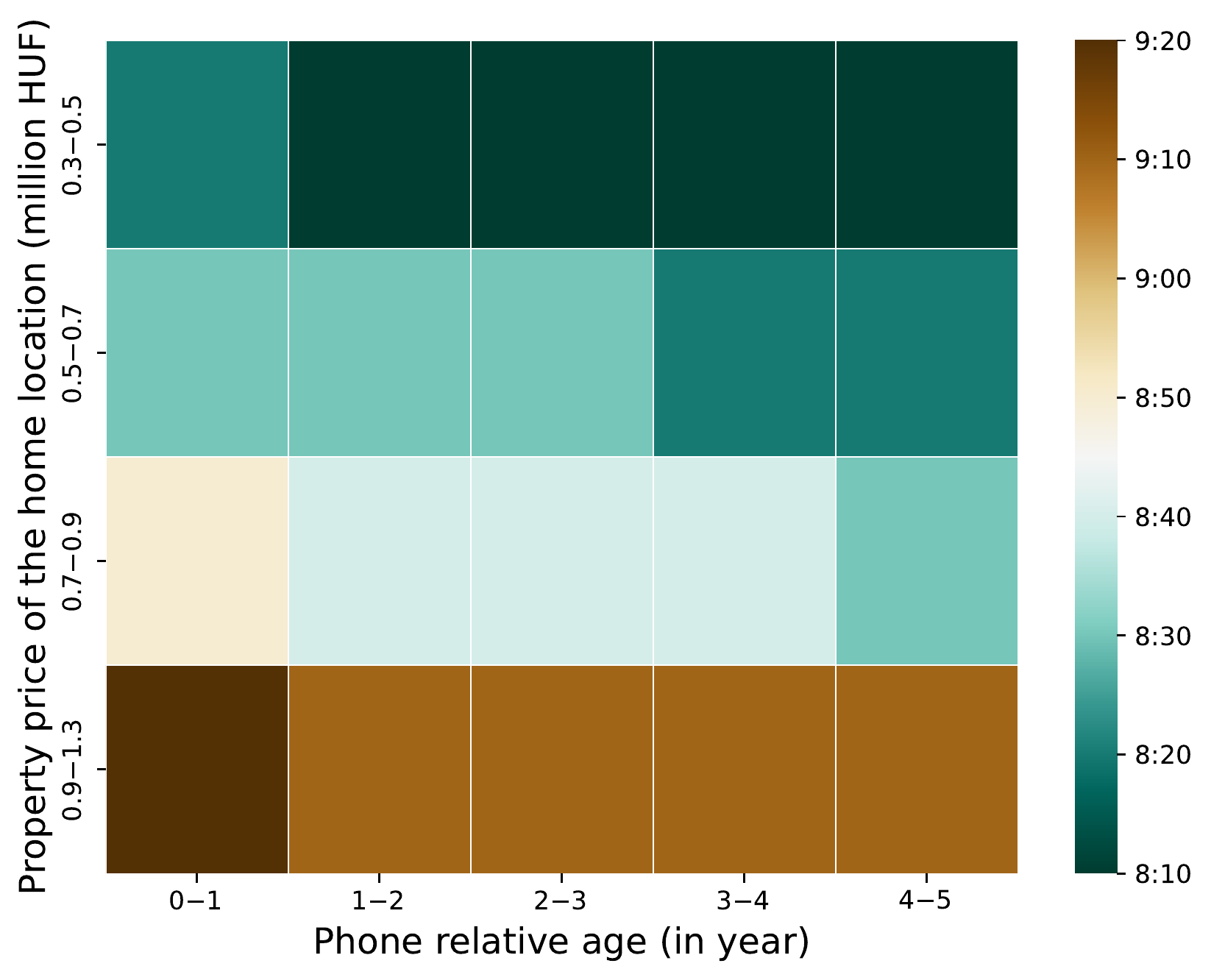}
        \captionsetup{position=bottom,justification=centering}
        \caption{}
        \label{fig:hp_pa_wuwd}
    \end{subfigure}
    \hfill
    \begin{subfigure}[t]{0.4975\linewidth}
      \includegraphics[width=1\linewidth]{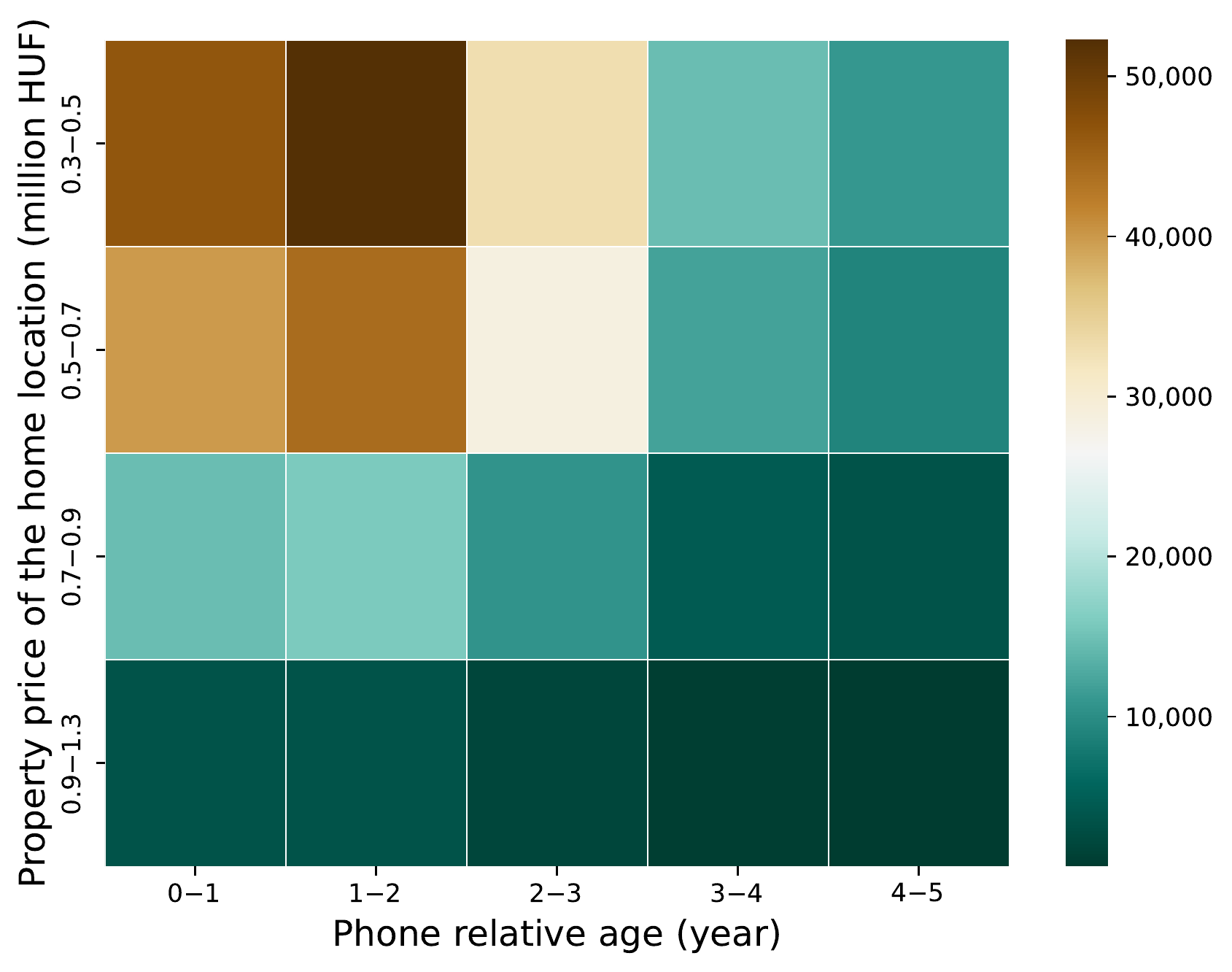}
      \captionsetup{position=bottom,justification=centering}
      \caption{}
      \label{fig:hp_pa_wuwd_cardinality}
    \end{subfigure}
    \caption{The inhabitant-based wake-up times in the socioeconomic categories, based on the property price of the home location and the relative age of the mobile phone (\textbf{a}), and the number of the subscribers in each category (\textbf{b}).}
    \label{fig:homeprice_phoneage_wakeup}
\end{figure}

As for Budapest, there is a positive correlation between the socioeconomic status and the wake-up time.
The inhabitant-based approach is used for this analysis, as the socioeconomic indicators are applied to subscribers.
Besides the real estate prices \cite{pinter2021evaluating}, two properties of the mobile phones \cite{pinter2021analyzing} were considered: the price and relative age. Figure~\ref{fig:homeprice_phoneprice_wakeup}a, shows the wake-up times in contrast of the real property price and the mobile phone price categories. To give context to these categories, Figure~\ref{fig:homeprice_phoneprice_wakeup}b illustrates the number of subscribers in each category.
Figure~\ref{fig:homeprice_phoneage_wakeup}, has the same structure, but applying to the age of the cell phones.

Four property price categories and five phone price categories are formed.
As Figure~\ref{fig:property_and_phone_price}a, shows most of the price of one square meter in most of the real estate advertisements are under 0.6 million \acrshort{HUF}. The property price categories are: (i) 0.3--0.5, (ii) 0.5--0.7, (iii) 0.7--0.9 and (iv) 0.9--1.3 million HUF.
Mobile phone price categories are: (i) 0--150, (ii) 150--300, (iii) 300-450, (iv) 450--600 and (v) 600--750 EUR. As for the mobile phone age, five categories were formed from 0 to 5 years. The older phones were omitted.
As Figure~\ref{fig:property_and_phone_price}, indicates most of the subscribers lived in less expensive homes and used less expensive cell phones, that were 1 to 3 years old.

In Figure~\ref{fig:homeprice_phoneprice_wakeup}a, the wake-up times are increasing in both dimensions. The lowest wake-up value is in the top-left corner, where the owners of the least expensive home locations and cell phones are grouped. Towards the bottom-right corner, the wake-up times are increasing.
This observation clearly indicates that, the richer people start their days somewhat later than the less wealthy people.
The differences are numerically not large between the categories, but the observation area is also relatively small. Budapest is 525 square kilometers, and its diameter is about 30 km. The farthest part of the agglomeration is about 40 km from the city center.

Figure~\ref{fig:homeprice_phoneage_wakeup}a, reveals negative correlation between phone age and the wake-up time. This result implies that wealthier people do not only use more expensive phones, but tend to use the latest models. Presumably, they replace their phones more often.

\subsection{Limitations}

As the activity curves do not reach their maxima immediately, the selected time does not accurately represent the neither the wake-up time, nor the time when the working hours actually start. These results can only be used relatively to other cells, or areas of the city. The reason of this, is that the \acrshort{CDR}s in this study only represents the active (in other word, billed) usage of the mobile phone network.
However, the employees usually do not actively use the network at the moment they arrived to their workplace or home. There is a certain gap between that moment and the first activity.
If the available data also contained the passive (cell-switching) communication, the time, when a \acrshort{SIM} card enter the home or work cell, could be used.
In this case, the terms ``leaving home'' and ``arriving to the workplace'' would be more accurate. Furthermore, the difference of the two could serve as a basis for a precise travel time estimation.

As the active mobile phone network activity is sporadic, the majority of the subscribers do not have enough activity records to trace back their daily movement accurately enough to determine when they leave their homes or arrive to their workplace. It is very common that, the subscribers do not have activity at home before work.
It would be necessary to have multiple activity in the morning to select the first and last home activity at home, to conclude when the subscriber tend to wake up and leave their home.
As this is not assured with this kind of data, this approach is not good for individual-level analysis.

The phone price database does not contain launch prices (see Appendix~\ref{app:iphone}) and the applied depreciation method is unknown. The results, utilizing the cell phone price as a socioeconomic indicator, should be interpreted by keeping that in mind.

\subsection{Future Work}

With the home and work locations, the distance can be determined, on the other hand, the travel time is hard to estimate. The subscriber could use different transportation modes, that have different time demands. Furthermore, without cell-switching information, the time when the subscriber left the home cell and arrived to the work call cannot be exactly determined.
Using the worker or inhabitant filtered activity curves --- described in Section~\ref{sec:working_hours} ---, the morning fall (and the evening rise) of the home location activity, and the rise (and the fall) of the work location activity can be calculated.
Considering that the work location activity rises when the workers are usually arrived, and the home activity of the inhabitants drops when they usually left to work, the difference of these values could be applied to estimate the travel time of a group of subscribers.
This would, naturally, require a larger number of subscribers in every home--work cell pairs, or the locations should be aggregated by base stations or even larger areas of the city.

Although, the ``edge detection'' method, used in this paper, gives reasonable results to determine the positive and negative edge of the activity curve, it would be worth to compare it with other approaches.

\section{Conclusions}
\label{sec:conclusions}

In this study, we introduced ``wake-up time'' as an indicator to describe the behavior of a group of subscribers. Due to the sporadic nature of the mobile phone network data, we grouped subscribers by home and work locations to analyze their activity at these locations.
By determining the rising and the falling edge of the activity curve of the workers at each location, the beginning, the ending, and the length of the working hours were estimated.
Tendencies between the starting and the ending time of the working hours at the work locations were also presented.
This indicator is used not only to classify the groups of subscribers (inhabitant-based approach), but the parts of Budapest (area-based approach), thus city parts can also be characterized by chronotypes. This was demonstrated by real-life examples as the opening hours of the malls, in Budapest, or by the late activity fall of the party district.

Wake-up time as a proposed indicator was compared to common indicators, as Radius of Gyration or Entropy, and clear negative correlation was found. On workdays, both Radius of Gyration and Entropy values were higher, while the wake-up times are lower. On holidays, quite the contrary. The correlation between bedtime, the counterpart of the wake-up time, and the mobility metrics was not that strong, but still considerable.

The day length, calculated as a difference of bedtime and wake-up time, was found to be constant between the workdays and holidays: the start and the end of the day is also shifted.
On the other hand, the day length reflected to the seasonal differences of the two data set: it was found that the days are longer, from the perspective a mobile phone network, when there is more daylight.
The longer days are the reason of the delayed activity fall, as the wake-up times found to be marginally affected by the earlier sunrise.

Using socioeconomic classes derived from housing prices at the home location, mobile phone prices and the age of the cell phone, correlation between the wake-up time and the socioeconomic status was also identified. The subscribers, living in less expensive apartments get up earlier, and this tendency holds true in respect of the mobile phones prices: subscribers, who owns more expensive cell phones tend to get up later.

These results may help to analyze further the city structures, by identifying ``early bird'' or ``night owl'' areas and possible connections between them. City parts with early morning or late night activities may require different public transport services, for example, and can aid the transportation infrastructure planning.
Business development could also benefit from the detailed insight of the neighborhood chronotypes, especially with the associated information of the home locations and the socioeconomic status of the subscribers.

The socioeconomic aspect of these findings can also contribute to better understanding the social structure of an urban environments.
In this regard, further studies need to be made, possibly with detailed census data, to evaluate the commuting and working habits of the different socioeconomic classes.

\vspace{6pt}

\section*{Author Contributions}
Conceptualization, G.P; methodology, G.P.; software, G.P.; validation, G.P.; formal analysis, G.P.; investigation, G.P.; resources, G.P. and I.F.; data curation, G.P.; writing---original draft preparation, G.P.; writing---review and editing, G.P.; visualization, G.P.; supervision, I.F.; project administration, I.F.; funding acquisition, I.F. All authors have read and agreed to the published version of the manuscript.

\section*{Funding}
This research supported by the project 2019-1.3.1-KK-2019-00007 and by the Eötvös Loránd Research Network Secretariat under grant agreement no. ELKH KÖ-40/2020.

\section*{Data Availability}
CDR and TAC data, used this study, are not publicly available due to third party restrictions.

\section*{Acknowledgments}
The authors would like to thank Vodafone Hungary and 51Degrees for providing the Call Detail Records and the Type Allocation Code database used in this study.
The estate price data are provided by the \url{ingatlan.com} estate selling portal.
For plotting the maps, OpenStreetMap data were used; these data are copyrighted by the OpenStreetMap contributors and licensed under the Open Data Commons Open Database License (ODbL).

\section*{Conflicts of Interest}
The authors declare no conflict of interest. The funders had no role in the design of the study; in the collection, analyses, or interpretation of data; in the writing of the manuscript, or in the decision to publish the results.

\printglossary[title=Abbreviations, toctitle=Abbreviations, nogroupskip=true]

\appendix
\section{Party District}
\label{app:party_district}

\begin{figure}
    \begin{subfigure}[t]{0.49\linewidth}
        \includegraphics[width=1\linewidth,trim={0 0 0 2.5cm},clip]{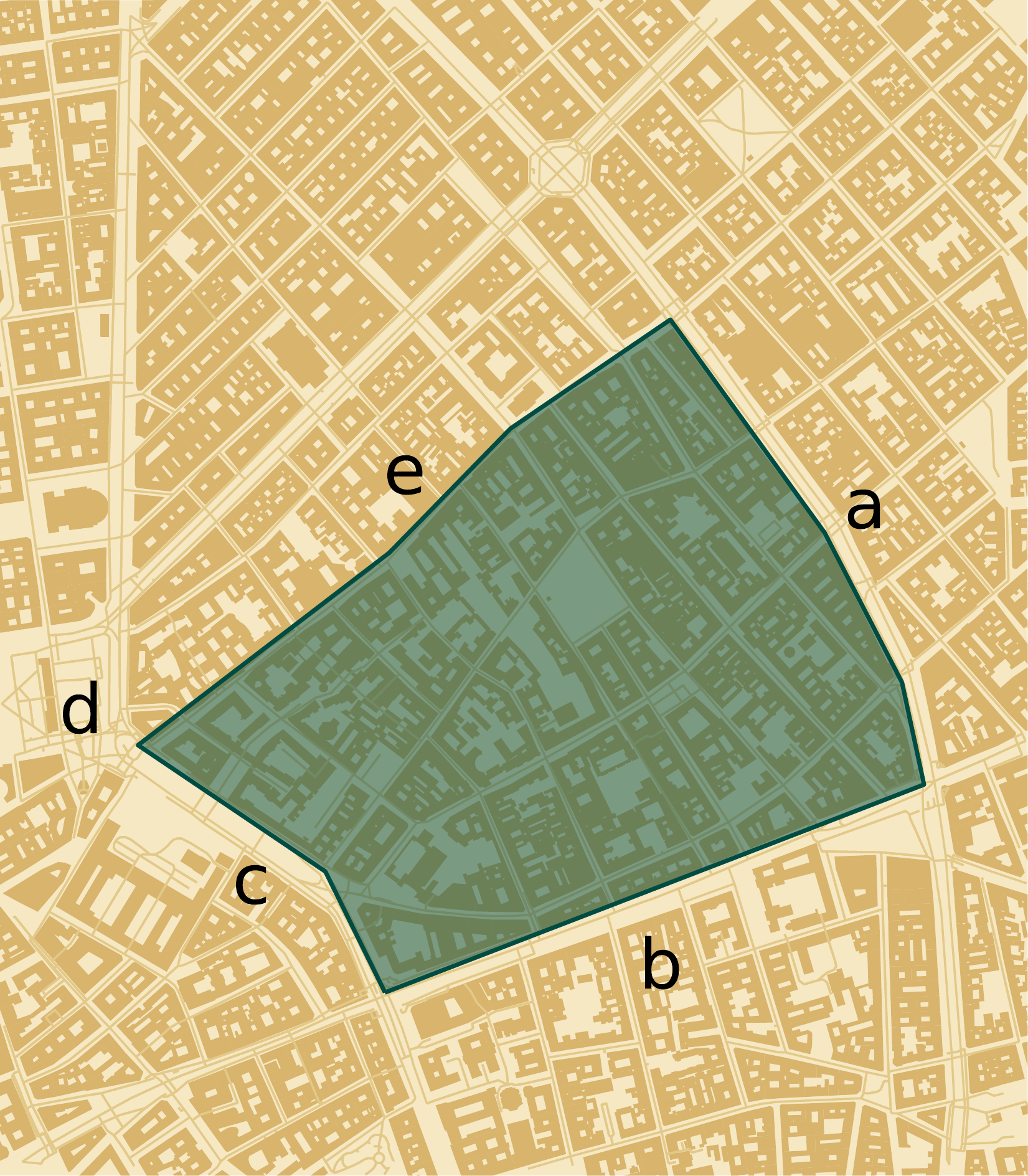}
        \captionsetup{position=bottom,justification=centering}
        \caption{}
        \label{fig:party_district}
    \end{subfigure}
    \hfill
    \begin{subfigure}[t]{0.49\linewidth}
      \includegraphics[width=1\linewidth]{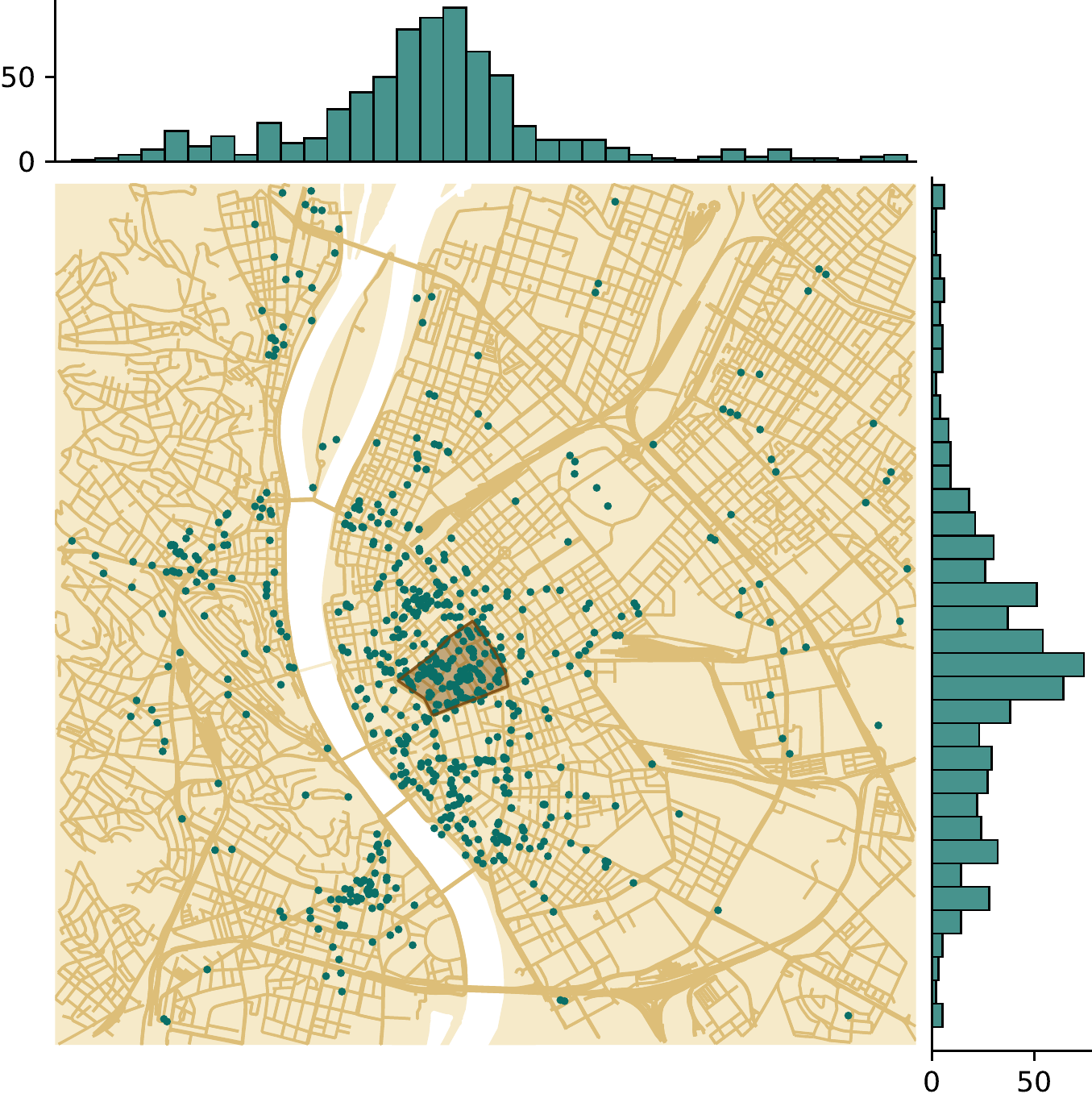}
      \captionsetup{position=bottom,justification=centering}
      \caption{}
      \label{fig:pubs}
    \end{subfigure}
    \caption{The borders of the party district (\textbf{a}), within District 7; and bars, pubs, beer gardens and nightclubs in Budapest downtown (\textbf{b}), based on OpenStreetMap data.}
    \label{fig:daily_gyr_ent_vs_wakeup}
\end{figure}

The so-called party district is not an official area of the city with definite borders. Usually, the area, bounded by the Károly Boulevard (c), Király Street (e), Erzsébet Boulevard (a) and Rákóczi Street (b) is referred as ``party district'' (see Figure~\ref{fig:party_district}).
This area is famous for ruin pubs and nightlife, which also involves that it is not the most active area early in the morning. Actually, the vivid nightlife of downtown is not confined only to this area. The Deák Ferenc Square (d), the northern side of the Király Street (e) and the eastern side of the Erzsébet Boulevard (a) can also be considered part of the party district.

Actually, how big is the party district? To answer this question, the points of interest (\acrshort{POI}) have been downloaded from the OpenStreetMap (\acrshort{OSM}) in the categories of ``bar'', ``pub'', ``biergarten'' and ``nightclub'', using the \acrshort{OSM} terminology, and plotted to a map (see Figure~\ref{fig:pubs}). There are 192 bars, 472 pubs, 12 beer gardens and 31 nightclubs in the displayed, 9~km by 9~km part of Budapest, and the area from Figure~\ref{fig:party_district} is also highlighted. Note that, the \acrshort{POI}s are queried in 2021, and cannot show the exact state of 2017. However, only the tendencies are important, that did not change fundamentally in the last decade.
The concentration of bars and pubs are, indeed, the highest within the highlighted ``party district'' area, but still very high on the whole Pest-side of the downtown, and there are also three smaller groups on the Buda-side of the city.

\section{iPhones}
\label{app:iphone}

\begin{figure}
    \begin{subfigure}[t]{0.5\linewidth}
      \includegraphics[width=1\linewidth]{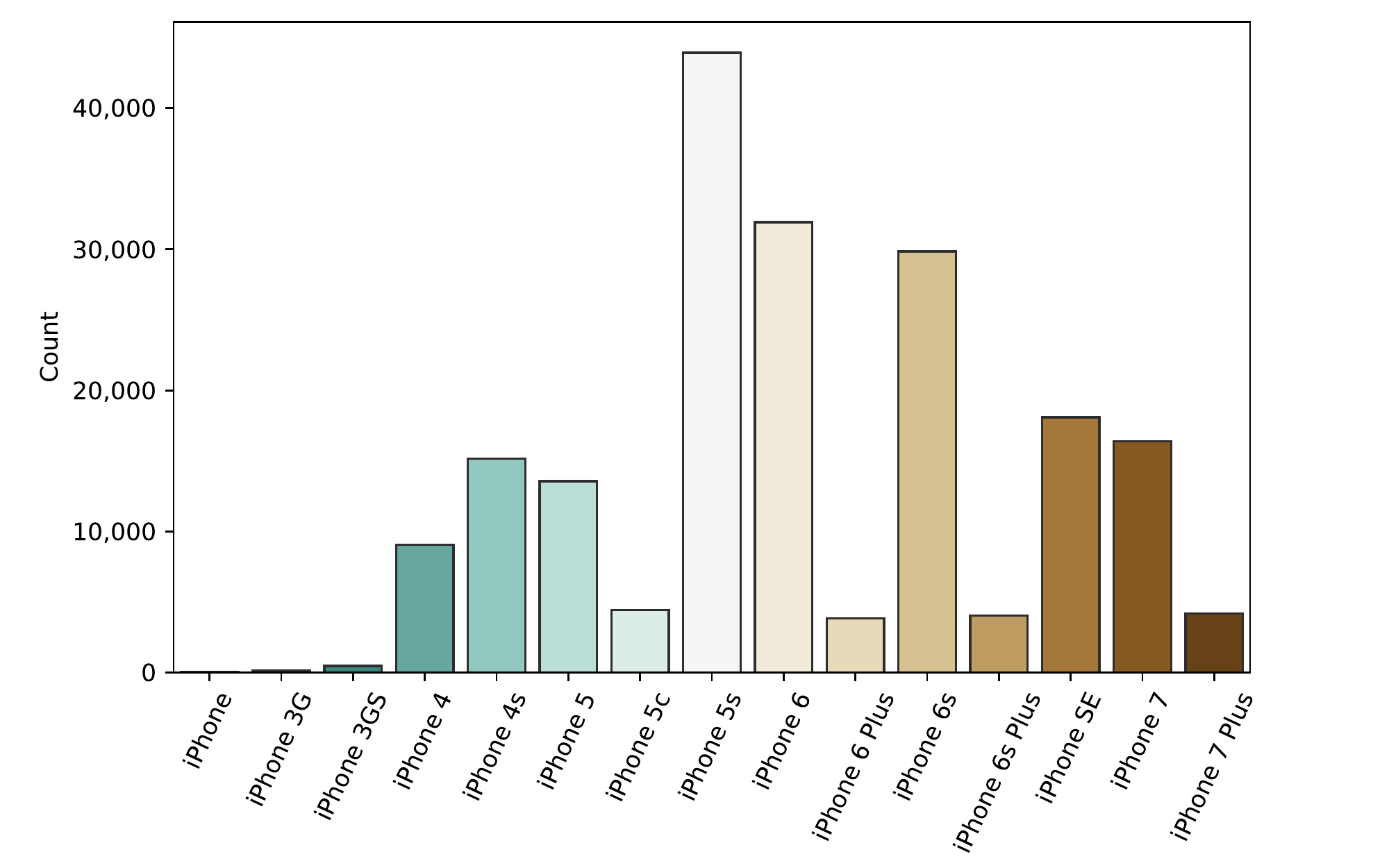}
      \captionsetup{position=bottom,justification=centering}
      \caption{}
      \label{fig:iphone_usage}
    \end{subfigure}
    \hfill
    \begin{subfigure}[t]{0.5\linewidth}
        \includegraphics[width=1\linewidth]{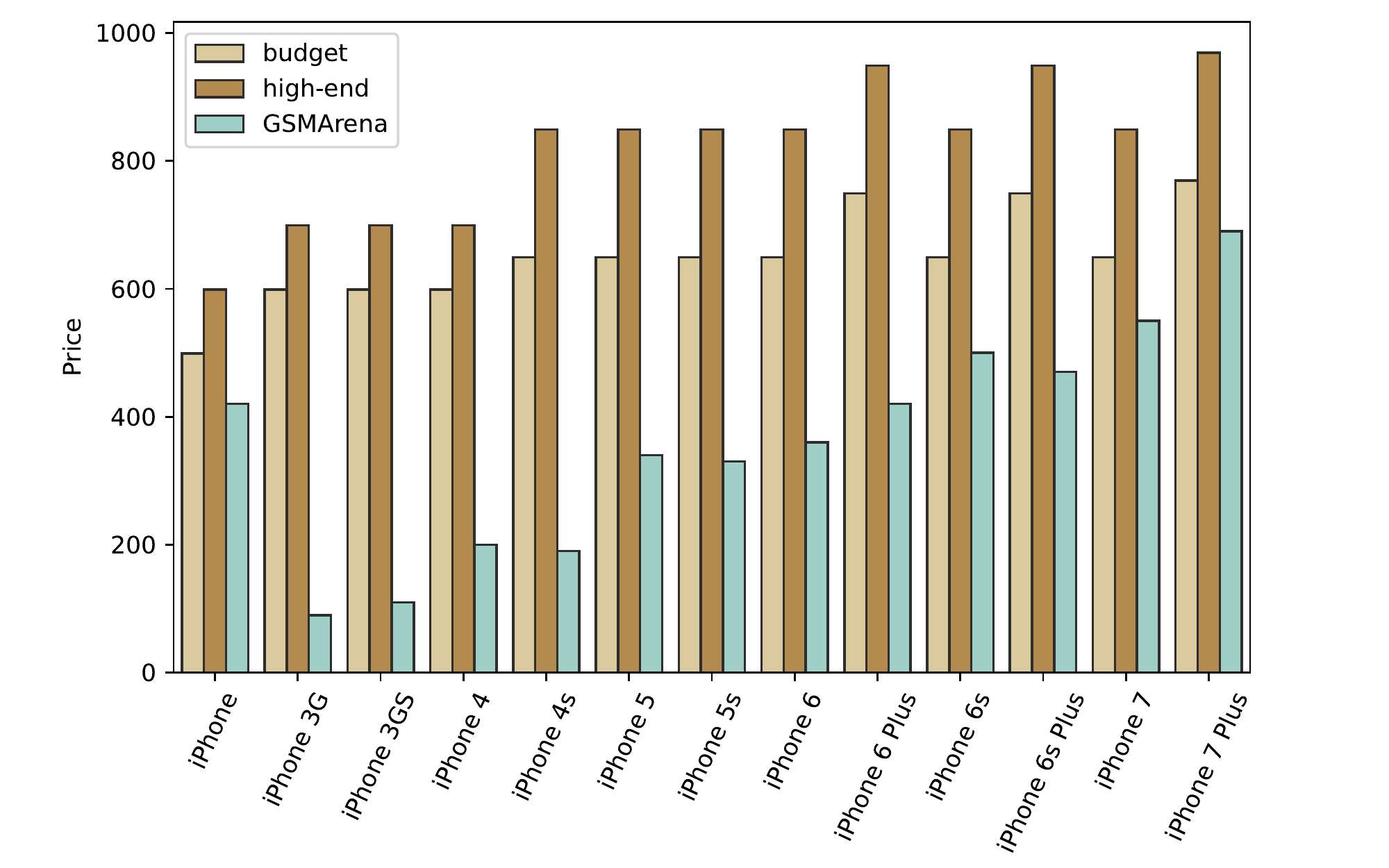}
        \captionsetup{position=bottom,justification=centering}
        \caption{}
        \label{fig:iphone_prices}
    \end{subfigure}
    \caption{
    Based on the ``April 2017'' dataset, the different iPhone models in use are also displayed (\textbf{a}), and
    comparing Apple iPhone prices \cite{protalinski2017iphone} with the GSMArena-based source \cite{mohit_gsmarena} (\textbf{b}).
    Versions with the lowest amount of storage denoted by ``budget'', and versions with the most expensive versions categorized as ``high-end''.
    }
    \label{fig:iphones}
\end{figure}

As Apple iPhones are considered a status symbol \cite{dissanayake2015role}, it makes them suitable to validate the phone price database \cite{mohit_gsmarena}, described in Section~\ref{sec:ses_data}. Figure~\ref{fig:iphone_usage}, shows the number of subscribers that exclusively use the certain iPhone models, in the ``April 2017'' dataset.
Using \acrshort{TAC} values, it is not possible to distinguish the iPhone models based on specification like storage. However, it is clear, that the most expensive models (``Plus'' versions) do not have significant user base, in contrast with some older models  like iPhone 4 and iPhone 5 series.

The launch prices of the iPhone models, released until April 2017, are obtained from \cite{protalinski2017iphone}.
Figure~\ref{fig:iphone_prices}, compares the two sources. As there are different versions of the certain models, a ``budget'' (with the lowest amount of storage), and a ``high-end'' (the most expensive) versions are also displayed.
Although, GSMArena price property is supposed to be a launch price, Figure~\ref{fig:iphone_prices}, clearly shows that they are much lower than the original prices. Moreover, the older the phone is, that lower the available prices are, except for the first iPhone.
It has to be noted, that GSMArena prices are in EUR, whereas the ground truth prices are in USD, but this cannot cause the difference. The results of this analysis implies, that the phone prices might have depreciated.

\printglossary[title=Abbreviations, toctitle=Abbreviations, nogroupskip=true]

\printbibliography

\end{document}